\documentclass[a4paper,fleqn,usenatbib]{mnras}

\usepackage[T1]{fontenc}
\usepackage{ae,aecompl}
\usepackage{verbatim}
\usepackage{wasysym}
\usepackage{pdflscape}
\usepackage{longtable}
\usepackage{lipsum}
\usepackage{float}
\usepackage{rotating, graphicx}
\let\oldAA\AA
\renewcommand{\AA}{\text{\normalfont\oldAA}}
\newcommand{\LX}{L_{\rm{X}}}

\usepackage{graphicx}	
\graphicspath{ {./images/} }
\usepackage{etoolbox}
\AtBeginEnvironment{longtable}{\singlespacing}
\usepackage{lineno}

\title[Correlations between X-ray properties and Black Hole
Mass in AGN]{Correlations between X-ray properties and Black Hole
Mass in AGN: a potential new method to estimate Black Hole Mass from short exposure X-ray observations}

\author[J.A Mayers et al]{Julian A. Mayers,$^{1}$ A. Kathy Romer,$^{1}$ Arya Farahi,$^{2,3}$ John P. Stott,$^{4}$ Paul Giles,$^{1}$
\newauthor Philip J. Rooney$^{1}$, A. Bermeo-Hernandez,$^{1}$, Chris A. Collins,$^{5}$ Matt Hilton,$^{6}$ Ben Hoyle,$^{7,8}$
\newauthor Andrew R. Liddle,$^{9}$ Robert G. Mann,$^{9}$ Christopher J. Miller,$^{2}$ Robert C. Nichol,$^{10}$
\newauthor Martin Sahl\'en,$^{11}$ C. Vergara-Cervantes,$^{1}$ Pedro T. P. Viana$^{12,13}$
\\
\\
$^{1}$Astronomy Centre, University of Sussex, Falmer, Brighton, BN1 9QH, UK\\
$^{2}$Astronomy Department, University of Michigan, Ann Arbor, MI 48109, USA\\
$^{3}$McWilliams Center for Cosmology, Department of Physics, Carnegie Mellon University, Pittsburgh, PA 15213, USA\\
$^{4}$Department of Physics, Lancaster University, Lancaster LA1 4YB, UK\\
$^{5}$Astrophysics Research Institute, Liverpool John Moores University, IC2, Liverpool Science Park, 146
Brownlow Hill, Liverpool, L3 5RF, UK\\
$^{6}$Astrophysics \& Cosmology Research Unit, School of Mathematics, Statistics \& Computer Science, University of KwaZulu-Natal, Durban 4041, SA\\
$^{7}$Universitaets-Sternwarte, Fakultaet fuer Physik, Ludwig-Maximilians Universitaet Muenchen, Scheinerstr. 1, D-81679 Muenchen, Germany\\
$^{8}$Max Planck Institute fuer Extraterrestrial Physics, Giessenbachstr. 1, D-85748 Garching, Germany \\
$^{9}$Institute for Astronomy, University of Edinburgh, Royal Observatory, Blackford Hill, Edinburgh, EH9 3HJ, UK\\
$^{10}$Institute of Cosmology and Gravitation, University of Portsmouth, Dennis Sciama Building, Portsmouth, PO1 3FX, UK\\
$^{11}$ Department of Physics and Astronomy, Uppsala University, SE-751 20 Uppsala, Sweden\\
$^{12}$~Instituto de Astrof\'{\i}sica e Ci\^{e}ncias do Espa\c{c}o, Universidade do Porto, CAUP, Rua das Estrelas, 4150-762 Porto, Portugal\\
$^{13}$~Departamento de F\'{\i}sica e Astronomia, Faculdade de Ci\^{e}ncias, Universidade do Porto, Rua do Campo Alegre, 687, 4169-007 Porto, Portugal\\
}

\date{Accepted XXX. Received YYY; in original form ZZZ}

\pubyear{2018}

\begin{document}
\label{firstpage}
\pagerange{\pageref{firstpage}--\pageref{lastpage}}
\maketitle

\begin{abstract}
The normalised excess variance ($\sigma^2_{\rm{NXS}}$) parameter can be used to measure the X-ray variability of active galactic nuclei (AGN). 

Several investigations have shown that $\sigma^2_{\rm{NXS}}$ has a strong anti-correlation with both X-ray luminosity ($L_{\rm{X}}$) and black hole mass ($M_{\rm{BH}}$). Therefore, $\sigma^2_{\rm{NXS}}$ can be used to estimate $M_{\rm{BH}}$ if other measurements are not available. Unfortunately, $\sigma^2_{\rm{NXS}}$ can only be measured from long duration X-ray observations (tens of kiloseconds). By comparison, the typical exposure time during forthcoming eROSITA all-sky survey is only a few hundred seconds. eROSITA will yield large numbers (up to 1 million) of $L_{\rm{X}}$ values for AGN, but few, if any, $\sigma^2_{\rm{NXS}}$ measurements.
In this study, we have investigated the possibility of using eROSITA $L_{\rm{X}}$ data to estimate $M_{\rm{BH}}$. For this, we have used XMM-Newton observations of a sample of AGN drawn from the XMM-Newton Cluster Survey (XCS).  Using 18 (11) AGN with $\sigma^2_{\rm{NXS}}$ measurements in 10ks (20ks) segments, we reconfirm the strong correlation between $\sigma^2_{\rm{NXS}}$ and $L_{\rm{X}}$ found by other authors. Using 30 AGN with spectrally determined $L_{\rm{X}}$ values and reverbation method determined $M_{\rm{BH}}$ values, we show that these quantities are also correlated. Using 154 AGN with spectrally determined $L_{\rm{X}}$ measurements, we find that $L_{\rm{X}}$ values estimated from eROSITA countrates will be robust. We conclude that although it may be possible to use $L_{\rm{X}}$ measurements from eROSITA to estimate $M_{\rm{BH}}$ values for large samples ($>$10$^{6}$) of AGN, further tests of the $M_{\rm{BH}}$ to $L_{\rm{X}}$ correlation are needed, especially at higher redshifts. 
\end{abstract}

\begin{keywords}
galaxies: active -- galaxies: nuclei -- X-rays: galaxies -- quasars: supermassive black holes
\end{keywords}



\section{Introduction}
\label{sec:intro}

The general consensus in Astronomy is that every large galaxy harbours a super-massive black hole (SMBH) with masses in a range $10^6$ to $10^9$M$\odot$, e.g. \citep{2005SSRv..116..523F}.  
About 10\,per cent of these are revealed, at any epoch, by an extremely bright active galactic nucleus (AGN), e.g.  \citep{Gandhi2005}, with bolometric luminosities ranging from $10^{42}$ to $10^{46}$ erg\,s$^{-1}$. 
$M_{\rm{BH}}$ and the stellar luminosity concentration parameter \citep{2001MNRAS.326..869T}. Studying the mechanisms underlying AGN activity and feedback is essential to improve our understanding of galaxy evolution \citep[e.g.][]{2007MNRAS.380..877S}. 
And, as numerical simulations of the co-evolution of blackholes and galaxies continue to improve \citep[e.g.][]{2015MNRAS.454..913D}, it is essential that larger samples of $M_{\rm{BH}}$ are gathered from observations. These measurements are needed in order to constrain structure formation models.\\

The most direct method to measure $M_{\rm{BH}}$ is via stellar velocity dispersions \citep[e.g.][]{2000ApJ...539L..13G,2005ApJ...620..744G}. However, the technique can only be used in the very local ($z<0.025$) Universe and when there is no AGN in the core: an AGN would be so bright as to obscure the starlight from the galactic bulge. Where there is an active core, the best alternative to measure $M_{\rm{BH}}$ is to use reverberation mapping \citep{Blandford1982}. This technique measures the delay between changes in the continuum emission from the hot gas in the accretion disk and the response to these changes in the broad emission lines in the optical, UV and near IR part of the spectrum.  However, this method, only applicable to Type 1 AGN (i.e. those with broad emission lines), is costly in terms of telescope time because it requires time resolved, high signal to noise, spectroscopy. Therefore, to date, only a few dozen successful measurements have been made. The largest combination of $M_{\rm{BH}}$ measurements from reverberation mapping   comprises of just 63 AGN \citep{Bentz2015}, which in turn draws on various other surveys including \cite{1998ApJ...501...82P}, \cite{2012ApJ...755...60G}, and \cite{2000ApJ...533..631K}.  \\

Although reverberation mapping is unlikely, at least in the near term, to deliver SMBH masses for large ($>$100) samples of AGN, it can be used to calibrate indirect methods that are less costly in terms of telescope time. For example, it has been used to calibrate a method that is based on the width of broad optical emission lines measured from single-epoch optical or UV spectroscopy (e.g. \citealt{2002ApJ...571..733V}). Reverberation mapping has also been used to calibrate $M_{\rm{BH}}$ estimation from X-ray variability, \cite[e.g.]{Ponti2012}.\\

Since the early days of X-ray astronomy it has been known that X-ray emission is by far the most important contributor to the overall luminosity of AGN \citep{Elvis1978}, and that this
%
X-ray emission demonstrates significant variability over periods of hours to days \citep[e.g.][]{1988MmSAI..59..239M,1988pnsb.conf..285P}. 
The short timescale of the variability implies that the X-ray emitting region is very compact - since variability is governed by light-crossing time - and hence located close to the SMBH. The most accurate method to quantify X-ray variability of an AGN is to look at the power spectral density function (PSD). Analysis from EXOSAT, \citep[e.g.][]{Lawrence1993}, and later RTXE, \citep[e.g.][]{2002MNRAS.332..231U}, showed that the PSD could be modeled by a powerlaw with slopes of $\Gamma\simeq-2$, flattening at some `break' frequency.
\cite{2006Natur.444..730M} demonstrated the PSD break increases proportionally with $M_{\rm{BH}}$. This was also confirmed by \cite{2007MNRAS.380..301K} who also showed that $M_{\rm{BH}}$ and break frequency were intimately related. \\

Detailed PSD analysis of light-curves from SMBH and black hole binaries (BHB) indicated that the emission engine powering both AGN and BHBs are the same, for although the X-ray variability timescales differ between AGN (a few hundred seconds and up) and BHBs (seconds or less), the power spectra are very similar. Hence the variability difference can be accounted for by the difference in the mass of the central object (e.g. \citealt{2002MNRAS.332..231U}, \citealt{2003ApJ...593...96M}).\\

Unfortunately, PSD analyses necessitate long (typically tens of kiloseconds) X-ray observations of individual AGN. This is because the lowest observable frequency scales as $t^{-1}$, where $t$ is the observation exposure time.  Due to the requirement of long X-ray exposures, the PSD method of estimating $M_{\rm{BH}}$ cannot be applied to large samples of AGN. 
An alternative way to define X-ray variability, that is significantly less costly in terms of X-ray telescope time, is the Normalised Excess Variance ($\sigma^2_{\rm{NXS}}$) \citep[e.g.][]{Nandra1997}. This parameter is given by

 \begin{equation}
    \sigma^2_{\rm{NXS}} = \frac{1}{\bar{x}^2}\bigg[\frac{1}{N-1}\sum_{i=1}^{N} (x_i - \bar{x})^2 - \frac{1}{N}\sum_{i=1}^{N} \sigma^2_i\bigg],
    	\label{eq:nxs}
\end{equation}

where $N$ is the number of time bins in the light-curve of the source, $\bar{x}$ is the mean count rate, $x_i$ is the count-rate in bin $i$ and $\sigma^2_i$ the error in count rate in bin $i$. A positive value of $\sigma^2_{\rm{NXS}}$ implies that intrinsic variability of the source dominates the measurement uncertainty (and vice versa). As shown by \cite{1997scma.conf..321V}, the $\sigma^2_{\rm{NXS}}$ is simply the integral of the PSD over a frequency interval $\nu_{min}$ to $\nu_{max}$ i.e:

\begin{equation}
    \sigma^2_{\rm{NXS}} = \int_{\nu_{min}}^{\nu_{max}} P(\nu) d\nu,
\end{equation}

where $\nu_{min}$ = $T^{-1}$, $\nu_{max}$ = ${2\Delta T}^{-1}$, $T$ is the duration of the observation, and  $\Delta T$ is length of light-curve time bin. \\

The Poisson uncertainty on an individual measurement of $\sigma^2_{\rm{NXS}}$ has been been estimated by \cite{Vaughan2003} to be 

  \begin{equation}
    \Delta(\sigma^2_{\rm{NXS}}) = \sqrt {\bigg( \sqrt \frac{2}{N} . \frac{\overline{\sigma^2_{err}}}{\bar{x}^2} \bigg)^2 + \bigg( \sqrt \frac{{\overline{\sigma^2_{err}}}}{N} . \frac{2\sigma_{nxs}}{\bar{x}}
\bigg)}
   	\label{eq:nxs_err}
\end{equation}

where $\overline{\sigma^2_{err}}$ is the mean square error. \\

When comparing different AGN, the $\sigma^2_{\rm{NXS}}$ values need to be $k$-corrected and adjusted to account for differences in observing times, see \cite{2013ApJ...779..187K}. According to \cite{2016JPhCS.689a2006M} these two factors can be accounted for by the scaling relation:

  \begin{equation}
    {^\star \sigma^2_{\rm{NXS}}}= \sigma^2_{\rm{NXS}} \bigg( \frac{\Delta t^{\star}}{\Delta t_{obs}} \bigg) ^{2\beta} (1 + z )^{2\beta},
   	\label{eq:nxs_scaling}
\end{equation}

where $z$ is the redshift of the AGN, $\Delta t^{\star}$ is a fixed time interval,  ${\Delta t_{obs}}$ is the time interval over the observation, and $\beta$ is estimated to be $0.10 \pm 0.01$ (e.g. \citealt{Antonucci2014}).\\


The use of the relationship between $\sigma^2_{\rm{NXS}}$ and $M_{\rm{BH}}$ as a proxy for $M_{\rm{BH}}$ was first proposed by \cite{2004MNRAS.350L..26N}. To date, the most comprehensive study of this relationship can be found \cite{Ponti2012}. This study drew upon the ``Catalogue of AGN In the XMM Archive'' (or CAIXA) published in  \cite{2009A&A...495..421B}. CAIXA contains 168 radio-quiet AGN that were observed by XMM-Newton, 125 of which had independent measurements of $M_{\rm{BH}}$ (32 of which coming from reverberation mapping).  Regardless of the $M_{\rm{BH}}$ estimation techniques, \cite{Ponti2012} found a significant anti-correlation between $M_{\rm{BH}}$ and $\sigma^2_{\rm{NXS}}$. This confirmed work by \cite{O'Neill2005} using ASCA observations of 46 AGN. \\

In addition to finding evidence of a relation between $M_{\rm{BH}}$ and $\sigma^2_{\rm{NXS}}$, \cite{Ponti2012} also reported a significant anti-correlation between $\sigma^2_{\rm{NXS}}$ and X-ray luminosity ($L_{\rm{X}}$). This confirmed measurements, albeit at lower significance by \cite{Lawrence1993}, \cite{Barr1986}, \cite{2009A&A...495..421B},  \cite{O'Neill2005}, \cite{Papadakis2004}, and  \cite{2010ApJ...710...16Z}.
Some authors (e.g \citealt[]{O'Neill2005, Papadakis2004}) have suggested that this anti-correlation is a bi-product of a `fundamental' relationship between  $M_{\rm{BH}}$ and $\sigma^2_{\rm{NXS}}$.\\




However, the results presented in \cite{Ponti2012} remain controversial. Subsequent work, \cite[e.g.][]{2015ApJ...808..163P} and \cite{2015MNRAS.447.2112L} -- based on the analysis of 11  and 14 low mass, $M_{\rm{BH}}$ $\leq 10^6M_{\odot}$, AGN respectively -- suggest that there is a flattening of correlation between variability and $M_{\rm{BH}}$ in the low mass regime. 
Moreover, \citealt{2013ApJ...771....9A},  conclude that $\sigma^2_{\rm{NXS}}$ is a \textit{biased} estimate of the variance of a continuously sampled light-curve which depends on both prior knowledge of the PSD slope and the sampling pattern. Moreover, the physical basis of a correlation between $\sigma^2_{\rm{NXS}}$ and $L_{\rm{X}}$ (and by implication between $M_{\rm{BH}}$ and $L_{\rm{X}}$) is unclear, as it would require the distribution of Eddington ratios (which can vary by up to three orders of magnitude for a given $M_{\rm{BH}}$, \citealt[e.g.][]{2002ApJ...579..530W}) to be peaked. Whereas some authors \cite[e.g.][]{2002ApJ...579..530W} claim there is no such peak, others claim that there is \cite[e.g.] {2010MNRAS.402.2637S}, \cite{2006ApJ...648..128K}, \cite{2012MNRAS.425..623L}.\\

The study presented here was motivated by evidence for a peaked distribution of AGN Eddington ratios, and by the forthcoming availability of up to a million $L_{\rm{X}}$ measurements for AGN from the upcoming eROSITA mission \citep{2010SPIE.7732E..0UP}.
Our goal was to determine whether it would be possible to gather useful $M_{\rm{BH}}$ estimates using $L_{\rm X}$ values alone. (Due to the short exposures of the individual observations comprising the eROSITA all-sky survey, very few -- if any -- values of $\sigma^2_{\rm{NXS}}$ will be measured.)
Our approach has been to re-examine correlations between the X-ray observables $L_{\rm X}$ and $\sigma^2_{\rm{NXS}}$, both with each other, and with reverberation mapping determined $M_{\rm{BH}}$ measurements. For the $\sigma^2_{\rm{NXS}}$ estimates, we have used the method advocated by \citealt{2013ApJ...771....9A}, which is more conservative than that used in \cite{Ponti2012}. An overview of the paper is as follows. 
Section\,\ref{Data} describes  how AGN were identified from the XMM Cluster Survey point source catalogue.  
Section\,\ref{NXS} describes methods used to the measure variability ($\sigma^2_{\rm{NXS}}$), X-ray luminosity ($L_{\rm{X}}$), and spectral index ($\Gamma$) of the AGN.
Section \ref{Scaling} presents the measured correlations between X-ray observables and $M_{\rm{BH}}$.
Section \ref{eROSITA} looks ahead to the launch of eROSITA and forecasts the potential to derive $M_{\rm{BH}}$ estimates from eROSITA luminosity measurements. 
Section \ref{Final} presents discussions and future directions. We assume the cosmological parameters $H_0=70{\rm{km\,s^{-1}\,Mpc^{-1}}}$, \ $\Omega_{\Lambda}=0.73$ and  $\Omega_{M}=0.27$ throughout.









\section{Data}
\label{Data}

The XMM Cluster Survey, (XCS)  \cite{2001ApJ...547..594R} provides an ideal opportunity to define a new sample of X-ray detected AGN. XCS is a serendipitous search for galaxy clusters using all publicly available data in the XMM-Newton Science Archive. In addition to collating detections of extended X-ray sources, i.e. cluster candidates, XCS also identifies serendipitous and targeted point-like X-ray sources. The XCS source catalogue grows with the size of the XMM public archive. At the time of writing, it contained over 250,000 point sources. The data reduction and source detection procedures used to generate the XCS source catalogue are described in \citep[][LD11 hereafter]{LloydDavies2011}. \\

For our study, we limited ourselves only to point-like sources detected by the XCS Automated Pipeline Algorithm ($\tt{XAPA}$) with more than 300 (background subtracted) soft-band photons. LD11 showed that, above that threshold, the XCS morphological classification (point-like versus extended) is robust. There are 12,532 such sources in the current version of the XCS point source catalogue. Using {\sc Topcat}\footnote{\tt http://www.star.bris.ac.uk/~mbt/topcat}, their positions have been compared to those of known AGN in VC13, and the SDSS-DR12Q Quasar Catalog \citep[][297,301 quasars]{2016arXiv160909489K}.\\


The matching radius was set to a conservative value of 5\,arcsec (LD11 find 95\,per cent of matches fall within within 6.6\,arcsec with a 1\,per cent chance of false identification within 10\,arcsec).  We find 2039 matches to XCS point sources ($> 300$ counts): 1,689 sources in VC13 and 513 in SDSS DR12Q, with 163 in common\footnote{Including XCS point sources with fewer than 300 counts, the number of matches increases to 6,505 sources in VC13 and 6,339 in SDSS DR12, with 890 sources in common.}. 
After removing sources within 10$^{\circ}$ of the Galactic plane, 1,316 remained sources in our sample (Sample-S0 hereafter, see Table\,\ref{table:AGNsamples}). \\



\begin{figure}
	\includegraphics[width=\columnwidth]{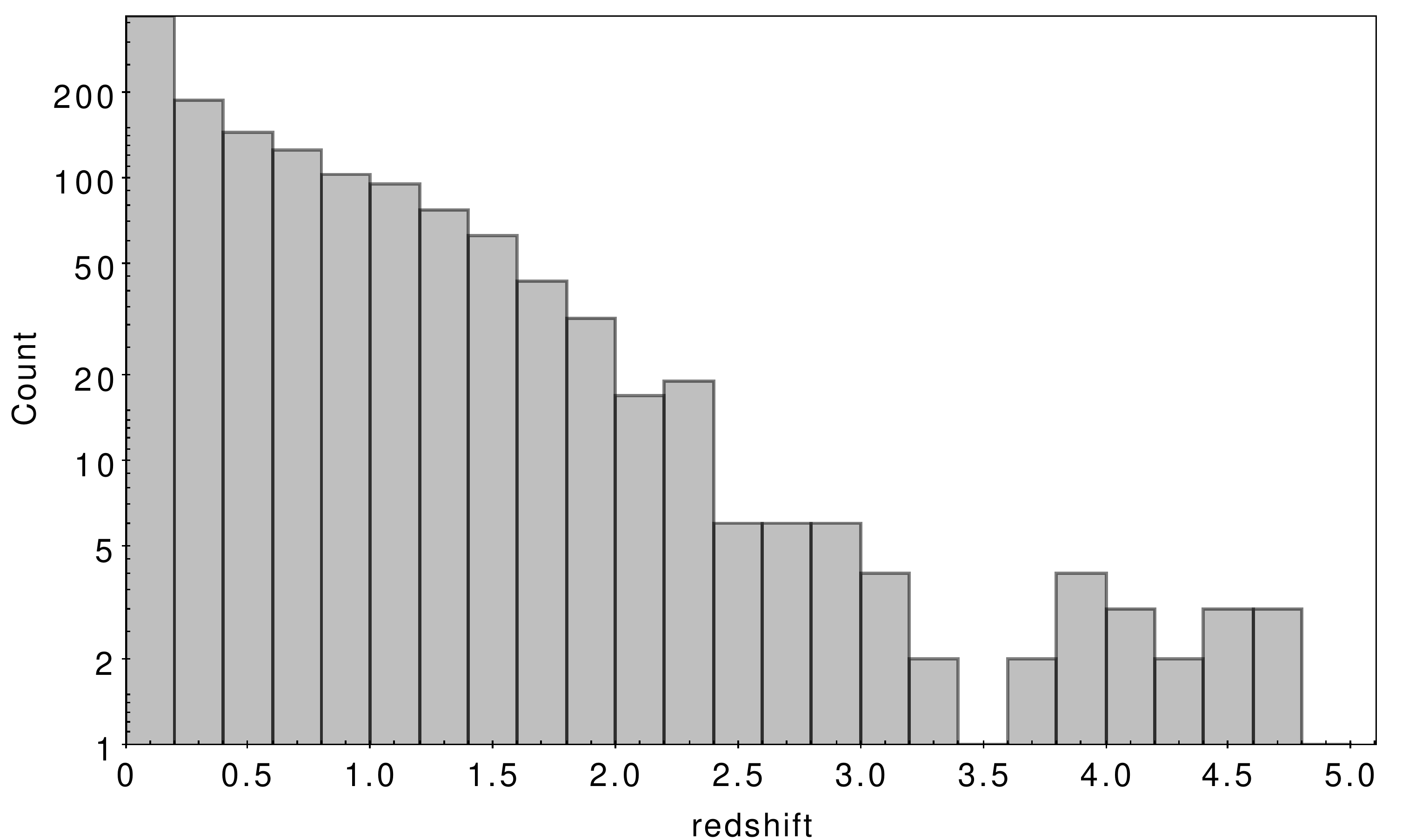}
	\includegraphics[width=\columnwidth]{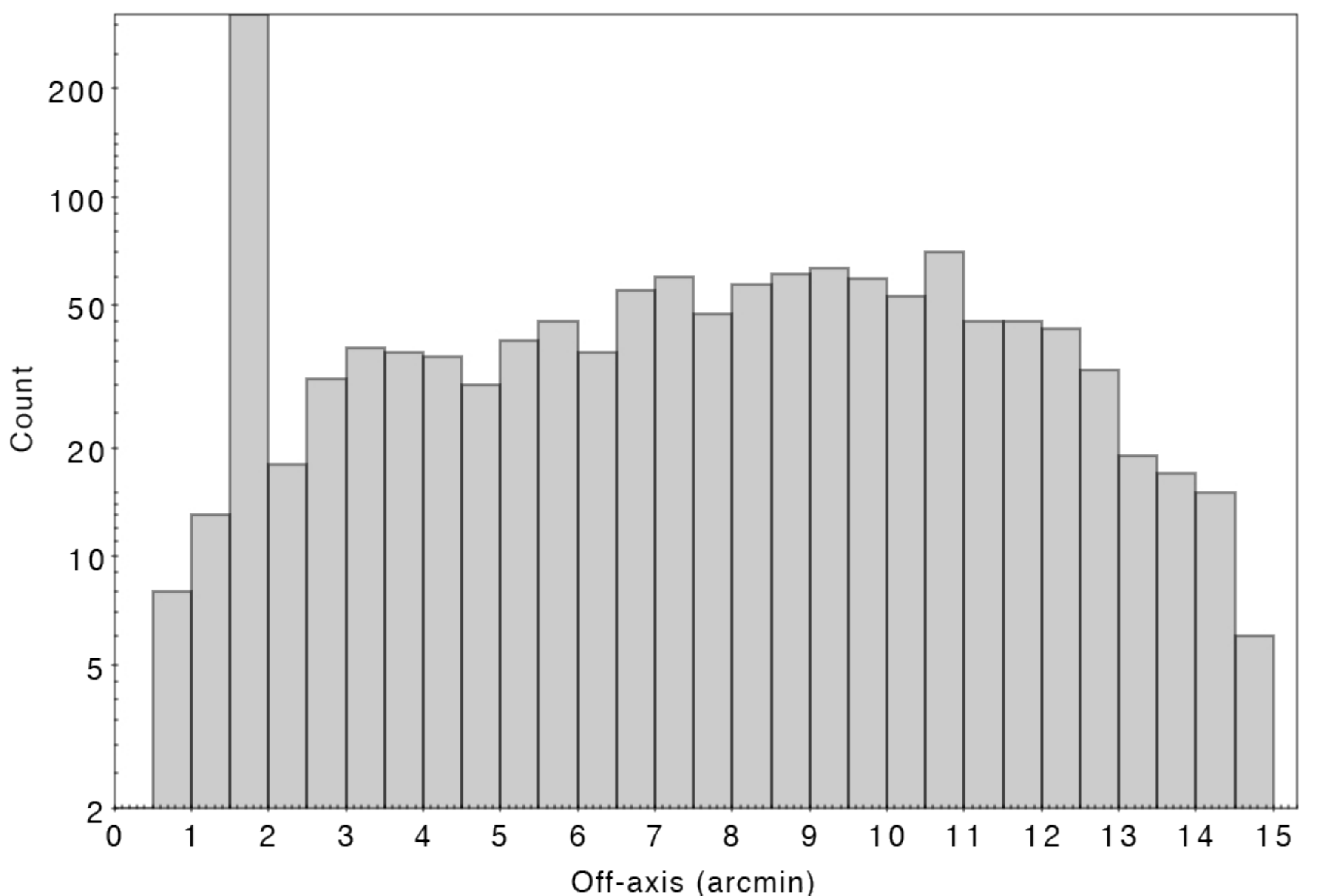}
	\includegraphics[width=\columnwidth]{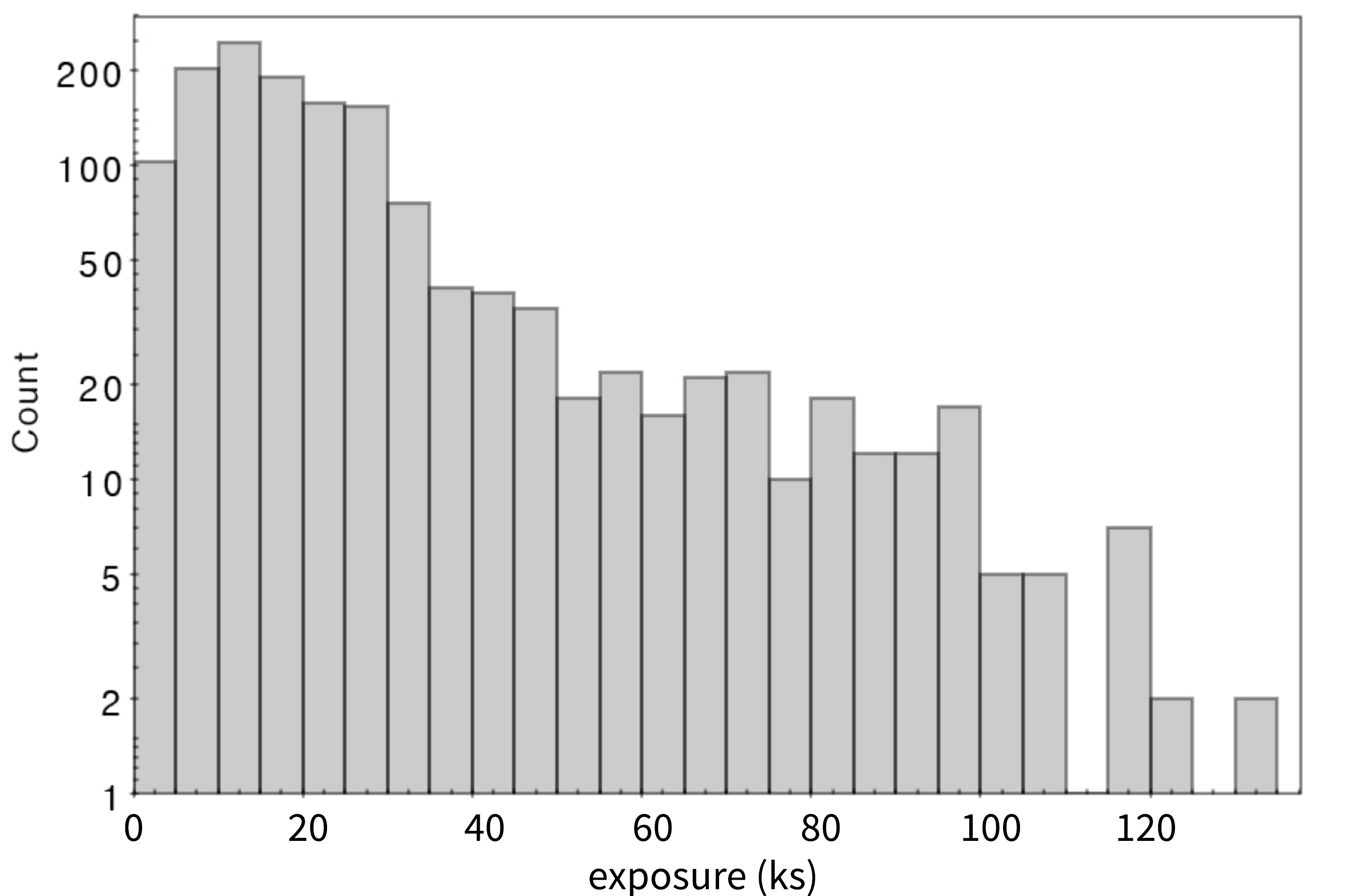}
    \caption{Properties of the 1,316 AGN in Sample-S0 (see Table\,\ref{table:AGNsamples}). Top: Redshift distribution.  Middle: Angular offset (arcmins) between the AGN location and the aim point of the respective XMM observation. Bottom: Full, i.e. before flare correction, XMM {\tt PN} camera exposure time of the respective XMM observation.}
    \label{fig:histos}
 \end{figure}

Redshifts for the AGN in Sample-S0 are taken from VC13 or from SDSS-DR12Q  - if the AGN appears in both catalogues, the VC13 value was used (note that there is minimal difference in redshift for AGN appearing in both catalogues). Many of the AGN were detected in multiple XMM observations. A total of 2,649 XMM observations have been included in the analyses of Sample-S0 presented herein.  The  distributions of redshift for each AGN, off-axis angle and the full observation duration  (i.e. before flare correction) of the 2,649 individual observations are shown in Figure\,\ref{fig:histos}.

\begin{center}
\begin{table*}
 \begin{tabular}{||c c c ||} 
 \hline
 Sample & Description & No.\\ [0.5ex] 
 (1)&(2)&(3)\\
 \hline\hline
 {\it Initial} & XCS sources in VC13 and/or SDSS-DR12Q & 2039$^1$ (46)\\ 
 S0 & Those in {\it Initial} with $b>10^{\circ}$  & 1316 (38)\\
 S1 & Those in S0 with $\frac{\Delta\LX}{\LX} < 1$ & 1091 (30) \\
 S10 & Those in S1 with $\sigma^2_{\rm{NXS}}$$_{[10ks]}$ measurements& 18 (8)  \\
 S20 & Those in S1 with $\sigma^2_{\rm{NXS}}$$_{[20{\rm ks}]}$ measurements & 10 (4)  \\


 \hline
\end{tabular}
\caption{Summary of the AGN samples used in the analysis. (1) sample name, (2) defines how the sample was filtered, (3) the number of AGN in the sample plus, in ( ), the number with masses from reverberation mapping. $^1$1689 from VC13, 513 from DSDSS-DR12Q with 163 common to both.}
\label{table:AGNsamples}
\end{table*}
\end{center}

\section{Data Reduction}
\label{NXS}
\subsection{Extracting light curves}
\label{sourceregions}


For each 
{\tt PN} observation of the 1,316 AGN in Sample-S0, we generate a clean event list which takes into account flare cleaning according to the methodology of LD11. We note that, for this study, we use only {\tt PN} detector data, because the other two EPIC camera detectors ($\tt{MOS1}$ and $\tt{MOS2}$) are less sensitive, especially in the soft, $0.5-2.0\rm{\,keV}$, energy band.



We extracted, from the clean event list, the source light-curve from a circular region 20\,arcsec in radius centred on the $\tt{XAPA}$ coordinates for each AGN. We extracted a background light-curve from a circular annulus, centred on the same coordinate, with inner and outer radii of 50\,arcsec and 60\,arcsec respectively. The three radii (20, 50, 60\,arcsec) were chosen so that photons from AGNs observed at large off-axis angles (where the PSF is extended and elongated compared to on-axis) do not extend into the background region. When other $\tt{XAPA}$ detected sources overlap with the source or background apertures, they were removed (`cheesed out') using 20\,arcsec radius circles. \\

\begin{figure}
	\includegraphics[width=\columnwidth]{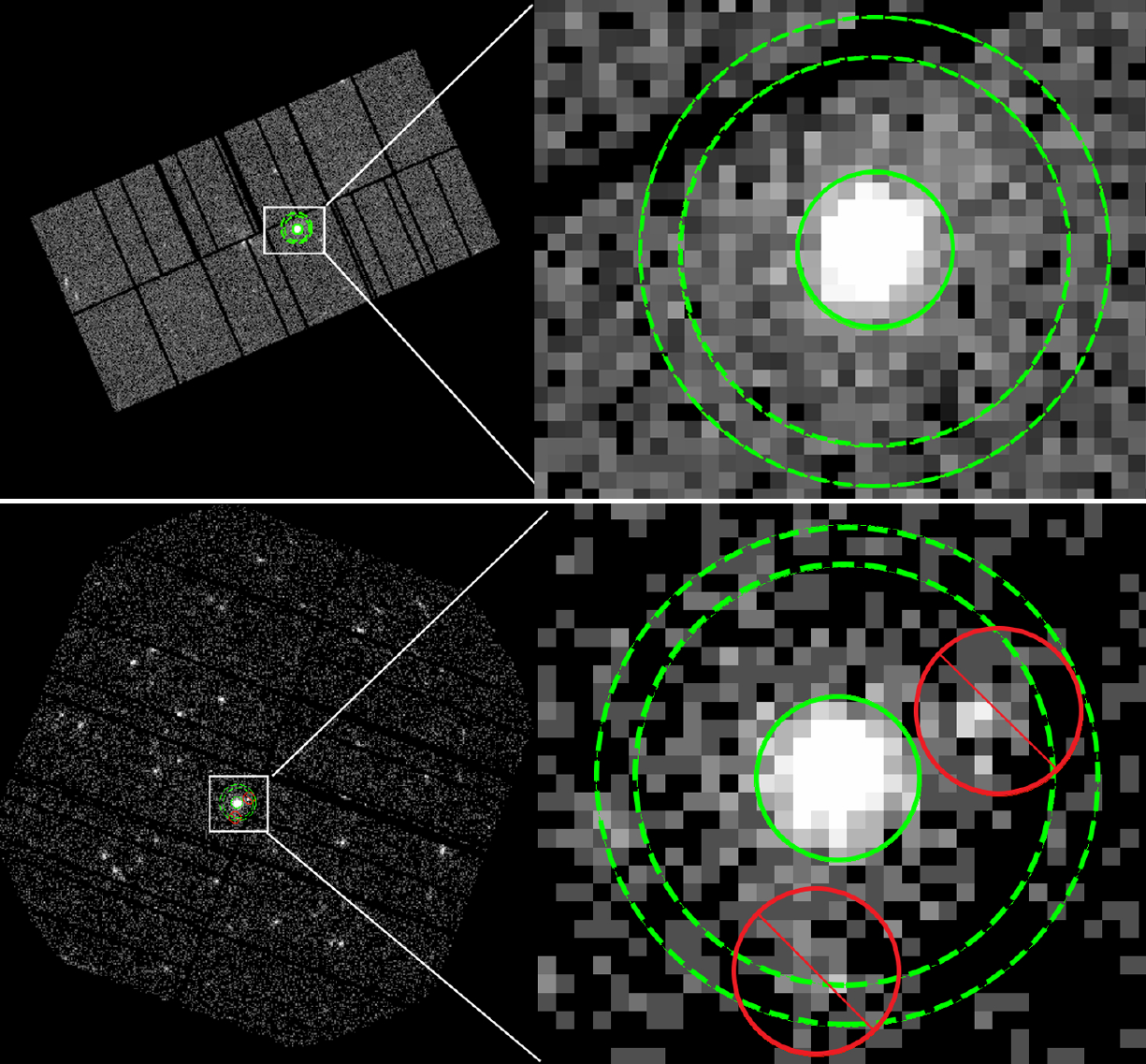}
	\includegraphics[width=\columnwidth]{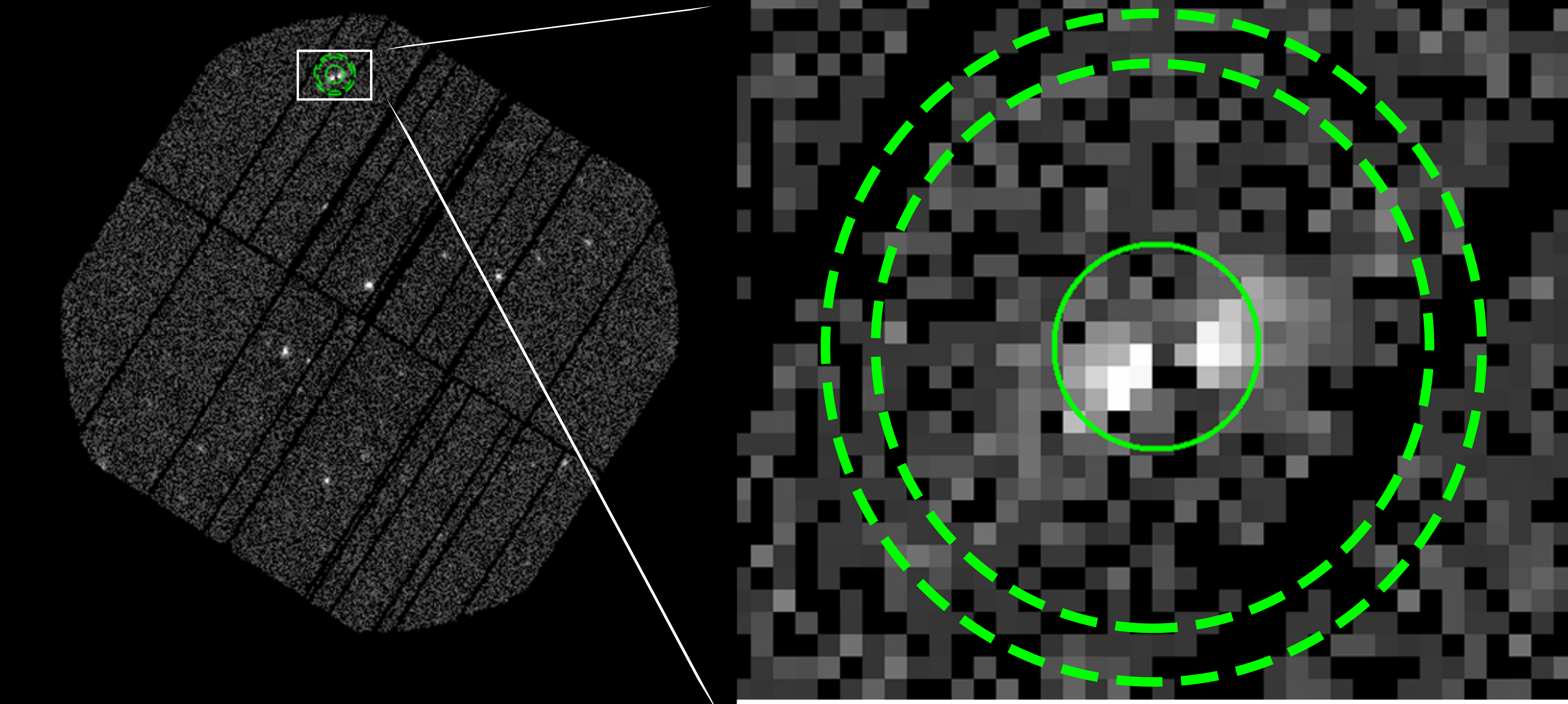}
    \caption{Top: $\tt{PN}$ ObsID 0673580301 and 16x zoomed in image of AGN XMMXCSJ132519.2-382455.2.  Middle: $\tt{PN}$ ObsID 0011830201 and $16 \times$ zoomed-in image of AGN  XMMXCSJ152553.9+513649.3. Areas around nearby point sources, outlined in red, are masked from the regions used to define light curves and spectra. Bottom: $\tt{PN}$ ObsID 0770190701 and $16 \times$ zoomed-in image of AGN  XMMXCSJ172034.0+580831.2 detected close to the edge of the field of view (14.3 arcmins off-axis). This AGN appears `bow tie' shaped due to the asymmetrical off-axis PSF. All: In each case the image were created with a pixel size of 4.52\,arcsec and within the energy range $2-10\rm{\,keV}$. The inner solid green circle (radius 20 arcsecs) defines the source region. The outer dashed dashed annulus (radii 50 - 60 arcsecs) defines the background region. }
    \label{fig:XMMXCSJ132519image}
\end{figure}

Figure\,\ref{fig:XMMXCSJ132519image} shows typical EPIC-{\tt PN} images of AGN in our sample. The source extraction regions are defined by the solid green circles, and background regions by the dashed green circles. The middle image shows an AGN with two nearby point sources (red circles). The bottom image shows an AGN detected close to the edge of the field of view. This sources shows the classic  `bow-tie' off-axis PSF shape as described in \citealt{LloydDavies2011}.\\

\begin{figure}
	\includegraphics[width=\columnwidth]{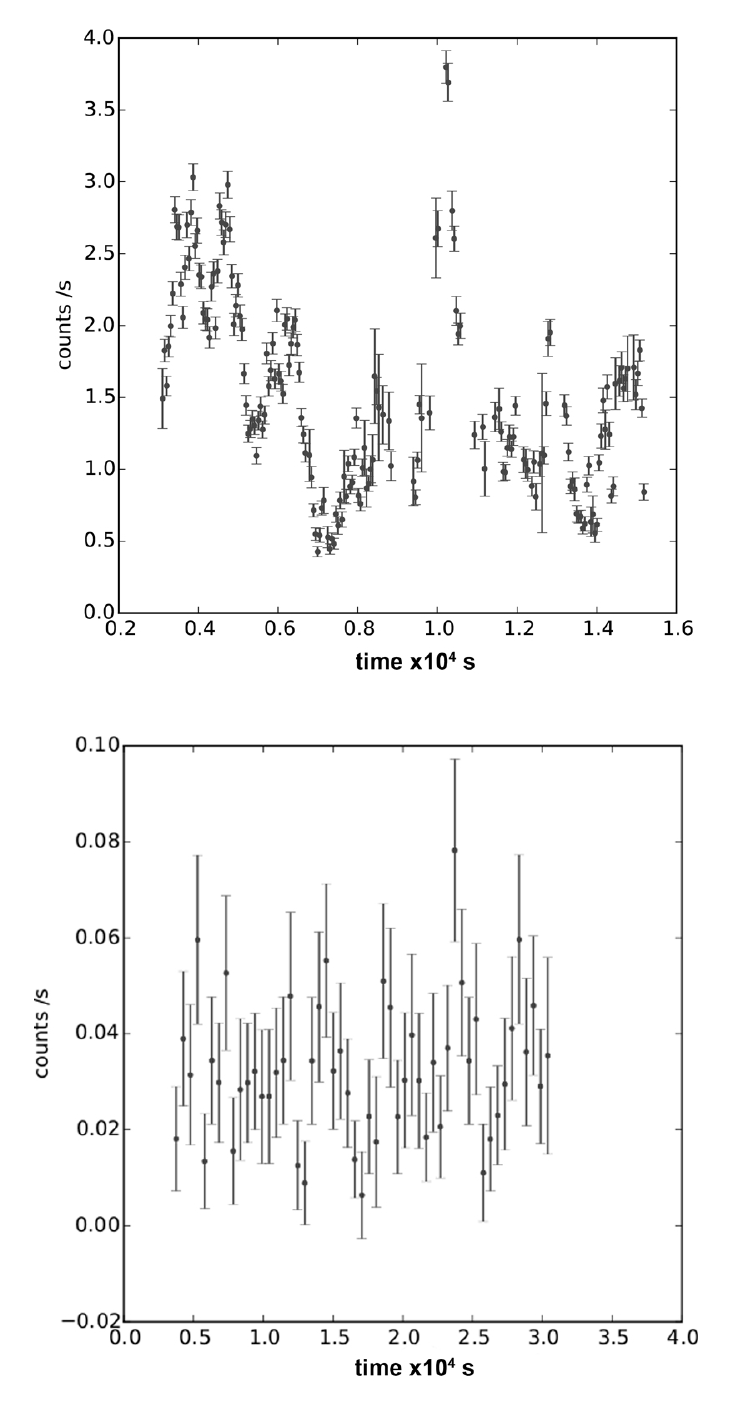}
    \caption{Background corrected light curves for AGN point-sources XMMXCSJ132519.2-382455.2 (top) and XMMXCSJ152553.9+513649.3 (bottom) taken from ObsID 0673580301 and 0011830201 respectively, with source and background regions extracted as shown in Figure\, \ref{fig:XMMXCSJ132519image} (top and middle). The binsize is 250 seconds and the energy range is $0.3 -10\rm{\,keV}$.}
    \label{fig:XMMXCSJ132519}
\end{figure}

We then generated rest frame $0.3-10.0\rm{\,keV}$ source and background light-curves in the $\tt{PN}$ detector  in 250s time bins using the XMM Science Analysis System ($\tt{SAS}$) task $\tt{EPICLCCORR}$ (this takes into account regions which fall on chip-gaps).  Figure\,\ref{fig:XMMXCSJ132519} shows the background corrected light curves for the AGNs featured in Figure\,\ref{fig:XMMXCSJ132519image}.

\subsection{Spectral fitting and luminosity estimates}


\label{sub:lum_estimate}
For each {\tt PN} observation of the 1,316 AGN in Sample-S0, we extracted, from the clean event lists, spectra for each AGN from the same source and background regions used to generate the light-curves (\S\,\ref{sourceregions}). 

The $\tt{arfgen}$ and $\tt{rmfgen}$ commands in $\tt{SAS}$ were used to generate the associated ancillary response files and detector matrices. The background-subtracted spectra were made such that there were a minimum of 20 counts in each channel. These were then were fit in the $0.2-10.0\rm{\,keV}$ energy range, to a typical AGN model \citep[e.g.][]{Kamizasa2012}: \begin{verbatim}phabs*cflux(powerlaw + bbody)\end{verbatim}  
using $\tt{XSPEC}$ v12.8.2.
The parameters in the {\tt cflux} model are {\tt Emin} and {\tt Emax} (the minimum/maximum energy over which flux is calculated), set to 0.001 and 100.0\,keV respectively, and {\tt lg10Flux} (log flux in erg/cm$^2$/s) which was left free. The other free parameters in our AGN model were the power-law index ($\Gamma$), black body temperature and black body normalisation. We fixed the value of $n_{\rm{H}}$ to the value taken from \citealt{Dickey1990}.
From the best fit model, we estimated the hard-band (from here defined as $2.0-10.0{\rm\,keV}$) luminosity, and its 68.3\,per cent confidence level upper and lower limits.\\

If the difference between the upper and lower limit on the luminosity (${\Delta\LX}$) was larger than the best fit value, i.e. $\frac{\Delta\LX}{\LX} > 1$ then the respective AGN was excluded from further analyses. The remaining sample contained 1,091 AGN 
and is referred to as Sample-S1 hereafter (see Table\,\ref{table:AGNsamples}). The median $\frac{\Delta\LX}{\LX}$ of this sample is 0.16.\\

\begin{figure}
	\includegraphics[width=\columnwidth]{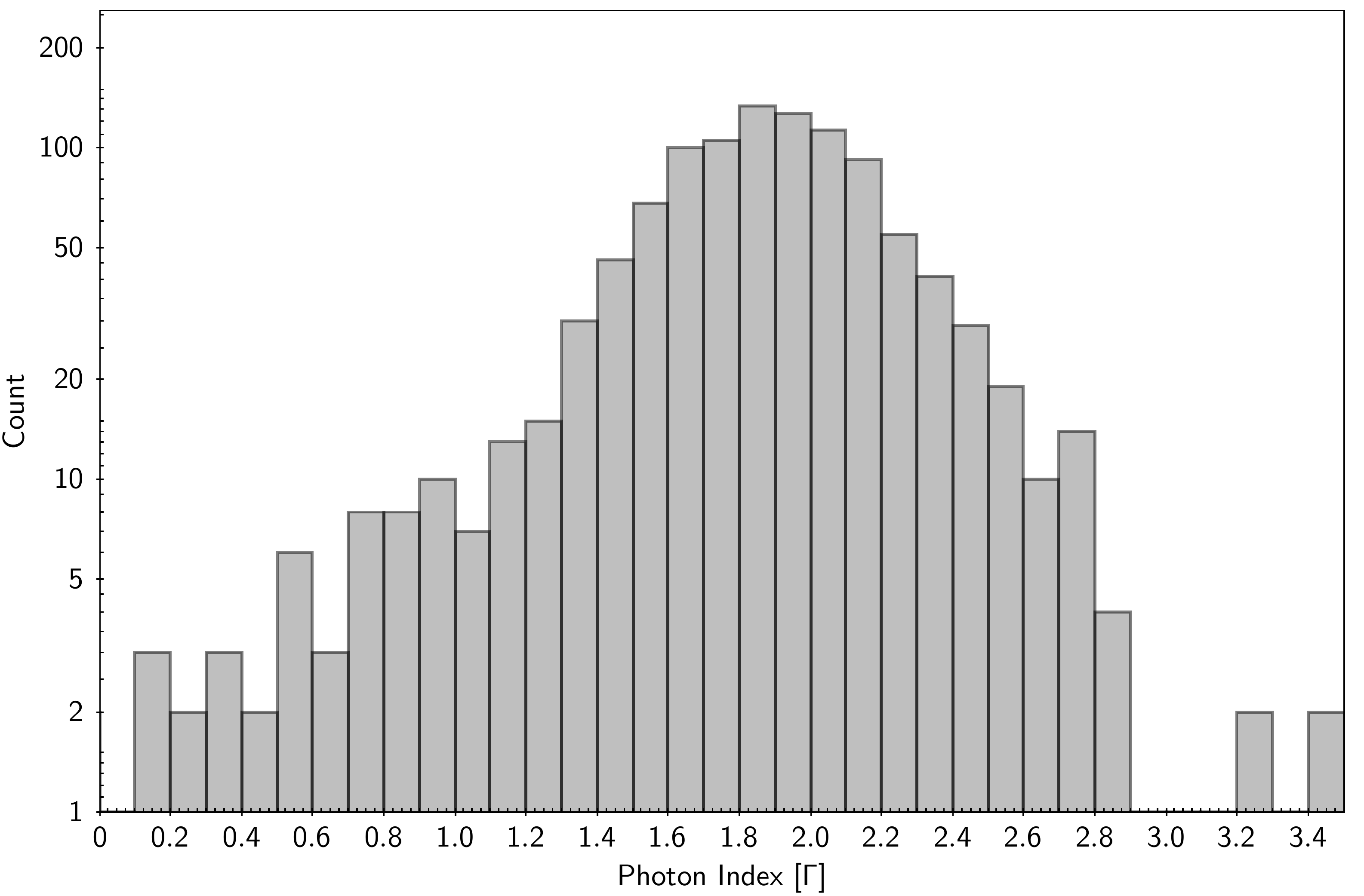}
	\includegraphics[width=\columnwidth]{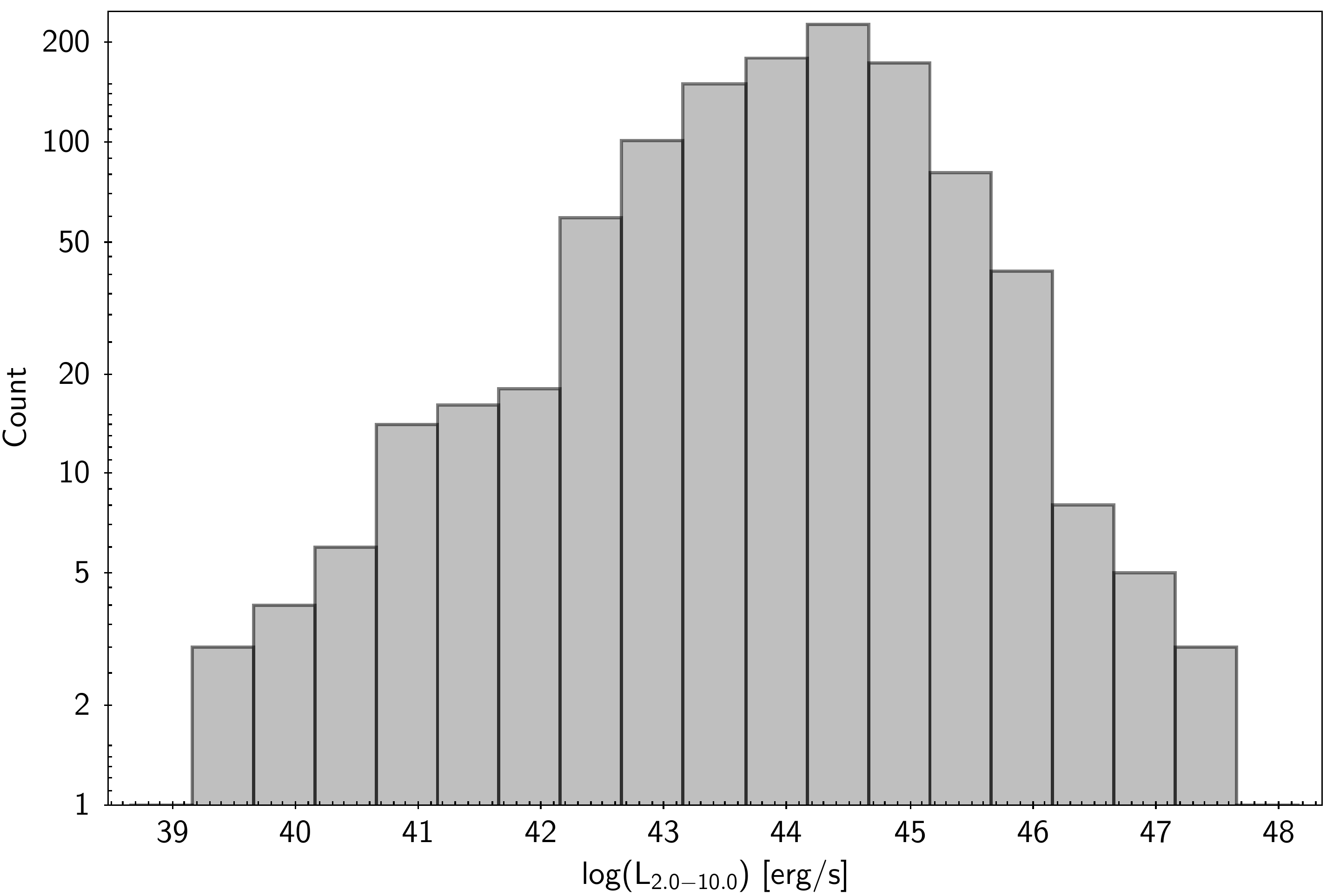}
    \caption{Histograms of the hard-band photon index $\Gamma$ (top plot) and hard-band luminosity (bottom plot) for Sample-S1 (1091 AGN).}
    \label{fig:histo_photonIndex}
\end{figure}

\begin{figure}
	\includegraphics[width=\columnwidth]{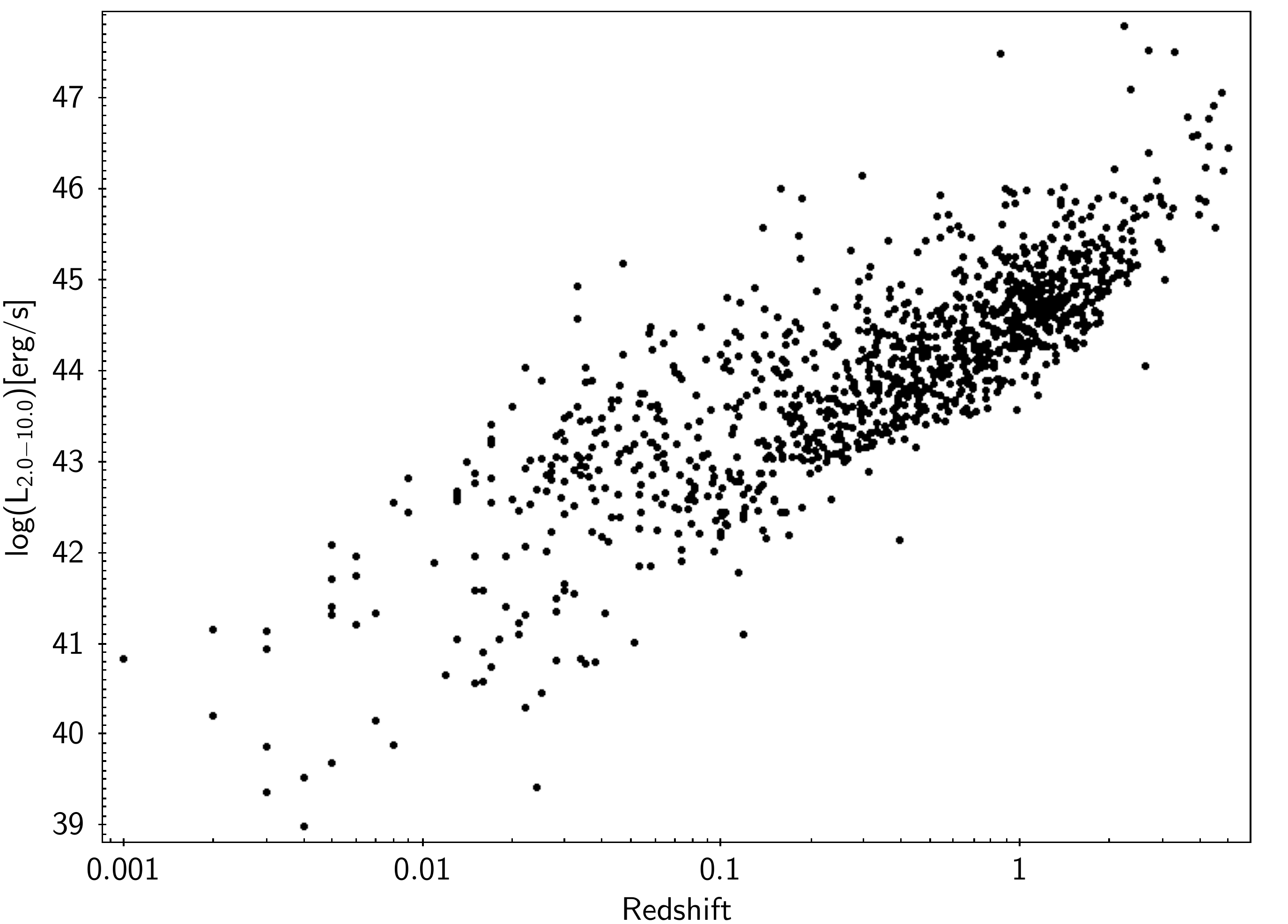}
    \caption{Distribution of Sample-S1 in the redshift-luminosity (hard-band) plane.}
    \label{fig:XCS_redshift_v_hLum}
\end{figure}

Figure\,\ref{fig:histo_photonIndex} (top), shows the distribution of power law index ($\Gamma$) and the bottom plot shows the hard-band luminosity for Sample-S1. Where there are multiple observations of the same source, we used the value derived from the observation with the longest on-axis exposure (in order not to underestimate luminosity). Figure\,\ref{fig:XCS_redshift_v_hLum} shows the redshift distribution versus hard-band-luminosity for Sample-S1.\\



The average hard-band spectral index was measured to be $\bar{\Gamma}=1.81\pm$0.34. This compares well with previous determinations. \citealt{Nandra1994} measured 1.9$<\Gamma<$2.0 using Ginga Large Area proportional Counter observations of  27 AGN.  \cite{2011A&A...530A..42C} found $\bar{\Gamma}=2.05\pm$0.03 using XMM observations of 305 AGN. \\

We note that in the cases that the AGN was detected at a large off-axis distance, our method will under-estimate the total $L_{\rm{X}}$ - because the extended PSF takes some of the source flux outside the 20\,arcsec aperture. Therefore, we stress that all of the AGN used to test $L_{\rm{X}}$ correlations in Section~\ref{Results} were detected on-axis. This missing flux issue does not effect the $\Gamma$ fit or $\sigma^2_{\rm{NXS}}$ measurement (Section~\ref{sec:rednoise}). 

\subsection{Determining normalised excess variance}
\label{sec:rednoise} 

The light-curves were divided into equal segments of 10\,ks (then again into segments of 20\,ks). Following the method of \citealt{O'Neill2005}, we required each time bin (within the segment) to have a minimum of 20 `corrected' counts - a value dependent on both the count-rate of that bin and the effective fractional exposure (EFE) of that bin. The EFE corrects for effects such as chip gaps and vignetting. Its value is stored in the $\tt{FRACEXP}$ parameter in the header of the background subtracted light-curve \textit{.fits} file. Also following \citealt{O'Neill2005}, any bin with $\tt{FRACEXP}<0.35$ was rejected if the corrected counts was $< 20$ (as in the case of a bright source with underexposed bins).  If after removing bins in this way there were less than 20 bins remaining, we reject the segment from our analysis. All remaining segments after these cuts are applied were labelled as \textit{good}.




The $\sigma^2_{\rm{NXS}}$ values for each AGN were then calculated using Equation\,\ref{eq:nxs} and a simplified version of  Equation\,\ref{eq:nxs_scaling}, i.e.

  \begin{equation}
    {^\star \sigma^2_{\rm{NXS}}}= \sigma^2_{\rm{NXS}}  (1 + z )^{2\beta}
   	\label{eq:nxs_scaling_simple}
\end{equation}

This simplification is possible because we measured $\sigma^2_{\rm{NXS}}$ in light-curve segments of consistent lengths, i.e. the $({\Delta t^{\star}}/{\Delta t_{obs})^{2\beta}}$ term in Equation\,\ref{eq:nxs_scaling} is not needed. Hereafter we still use the term $\sigma^2_{\rm{NXS}}$, rather than $^\star \sigma^2_{\rm{NXS}}$ to describe normalized excess variance even after the $(1 + z )^{2\beta}$ correction has been applied. 


For each AGN we derived an unweighted mean value for $\sigma^2_{\rm{NXS}}$ from all \textit{good} 10\,ks and 20\,ks) light-curve segments. We calculated the $1\sigma$ error on this mean value as\footnote{(Equation 13 in \citealt{2013ApJ...771....9A}.)}

  \begin{equation}
    \Delta \sigma^2_{\rm{NXS}} = \sqrt {\sum_{i=1}^{n}  \Big| \sigma^2_{\rm NXS,i} - \overline{\sigma^2_{\rm{NXS}}} \Big| \Big/ \Big(n(n-1) \Big)}
   	\label{eq:nxs_av_error}
\end{equation}


where $n$ is the number of segments, $\sigma^2_{\rm{NXS},i}$ is the $\sigma^2_{\rm{NXS}}$ value of segment $i$ and $\overline{\sigma^2_{\rm{NXS}}}$ is the mean value across all \textit{n} segments.

\subsection{Mitigation of red-noise}

\begin{figure}
	\includegraphics[width=\columnwidth]{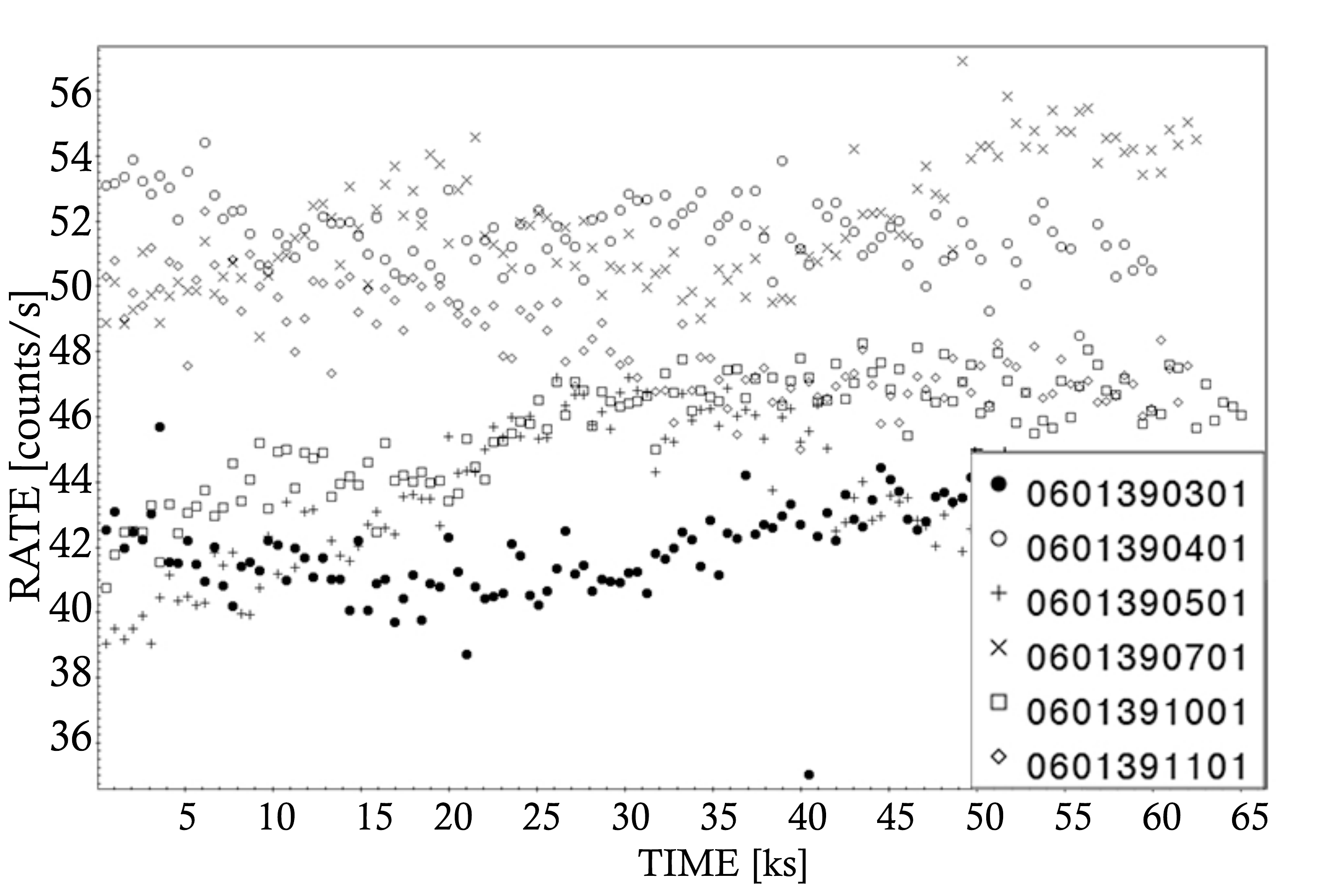}
    \caption{The $\tt{PN}$ detector light-curves in 250s bins of the AGN XMMXCSJ204409.7-104325.8 in ObsIDs 0601390301, 0601390401, 0601390501, 0601390701, 0601391001 and 0601391101 (0.5-10\rm{\,keV}).}
    \label{fig:XMMXCSJ204409_lightcurves_combined}
\end{figure}

\begin{figure}
	\includegraphics[width=\columnwidth]{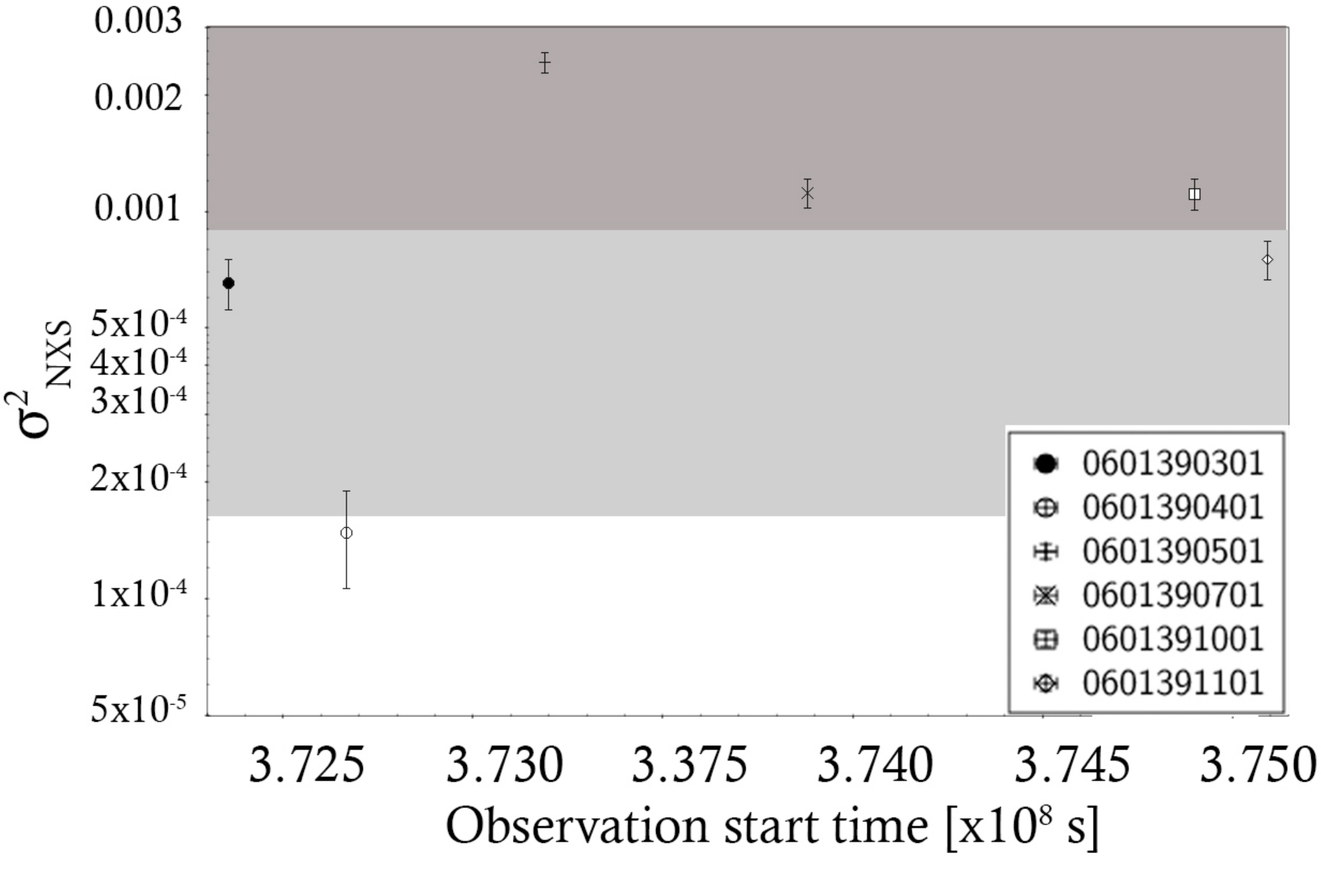}
    \caption{The measured $\sigma^2_{\rm{NXS}}$ value for the AGN XMMXCSJ204409.7-104325.8 using the lightcurves shown in Figure\,\ref{fig:XMMXCSJ204409_lightcurves_combined}. The respective observation start time is indicated on the x-axis. For reference, we show the 1 and 2$\sigma$ (dark and light grey respectively) confidence regions relative to the best fit $L_{\rm{X}}$-$\sigma^2_{\rm{NXS}}$ [20ks] relation at the mean $L_{\rm{X}}$ for this AGN (see Figure\,\ref{fig:XCS_hardband_lum_v_NXS_10ks_20ks_40ks_segs}).}
    \label{fig:XMMXCSJ204409_NXS_v_ObsTime_zoom_new_scatter}
\end{figure}

AGN exhibit a-periodic \textit{red-noise}, whereby there is an inherent uncertainty in the long-term variability due to the stochastic nature of AGN emission. As a result, an AGN light-curve generated at a given epoch is just one of many manifestations of the light-curve that the AGN will exhibit over its lifetime \citep[see][]{2003MNRAS.345.1271V}. Estimating \textit{red-noise} is difficult, as it depends on the steepness the of the PSD of the AGN. Thus, uncertainty regarding \textit{red-noise} would persist even if the measurement errors on a given $\sigma^2_{\rm{NXS}}$ could be reduced to zero. This is demonstrated in Figure\,\ref{fig:XMMXCSJ204409_lightcurves_combined}, which shows light curves for an AGN that was observed by XMM at multiple epochs, the offset in the normalization between the six curves demonstrates the underlying stochastic variability. Figure\,\ref{fig:XMMXCSJ204409_NXS_v_ObsTime_zoom_new_scatter} shows the respective $\sigma^2_{\rm{NXS}}$ value across the full duration of these observations, with the x-axis showing the start of the observation in XMM mission time. The dark and light grey shaded areas represent the 1 and 2$\sigma$ scatter regions relative to the best fit $L_{\rm{X}}$-$\sigma^2_{\rm{NXS}}$ [20ks] relation (as defined in Figure\,\ref{fig:XCS_hardband_lum_v_NXS_10ks_20ks_40ks_segs}) at the mean $L_{\rm{X}}$ of this AGN  (i.e. all the $\sigma^2_{\rm{NXS}}$ measurements are within 2$\sigma$ of the derived $L_{\rm{X}}$-$\sigma^2_{\rm{NXS}}$ relation.)

\begin{figure}
\includegraphics[scale=0.22]{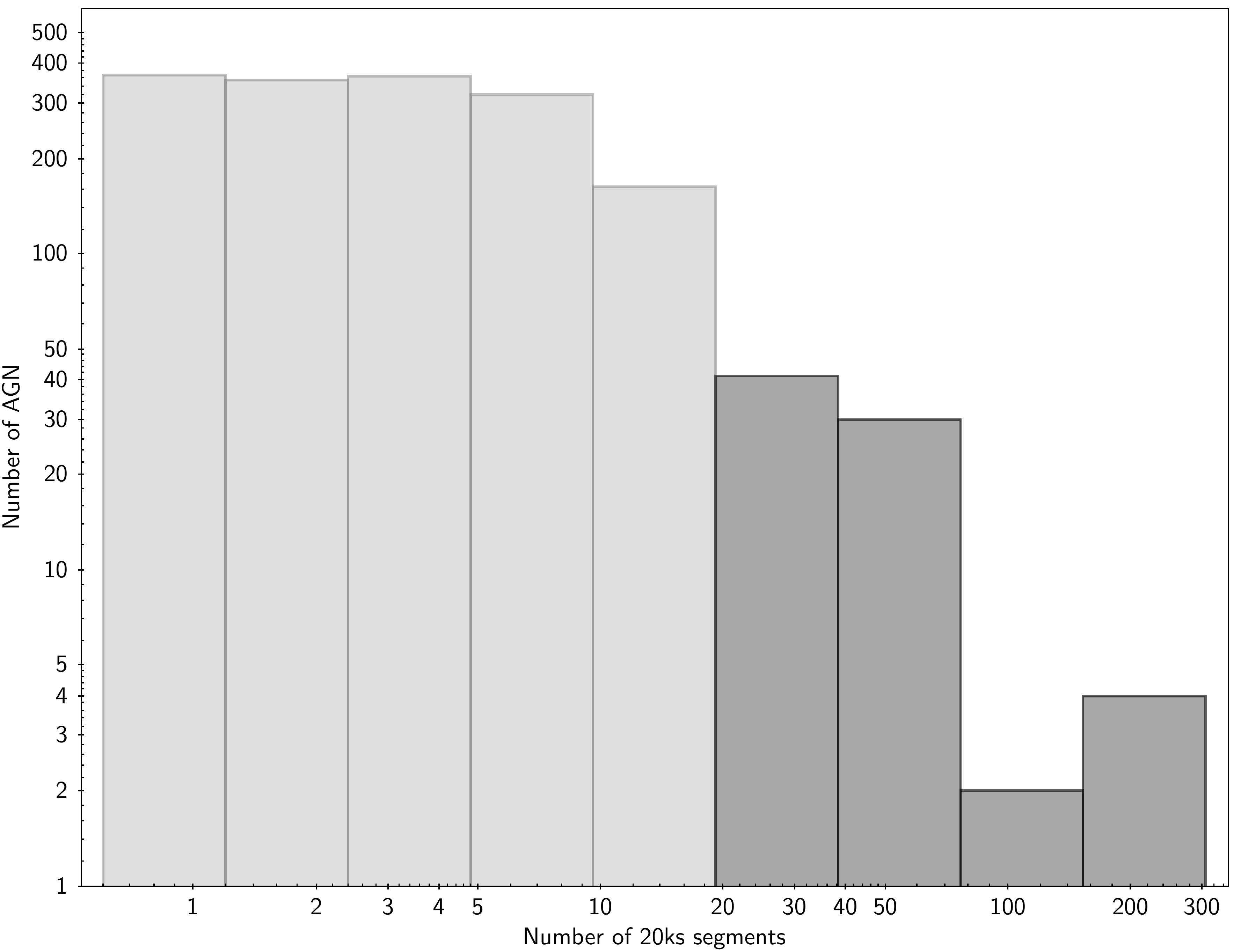}
\includegraphics[scale=0.22]{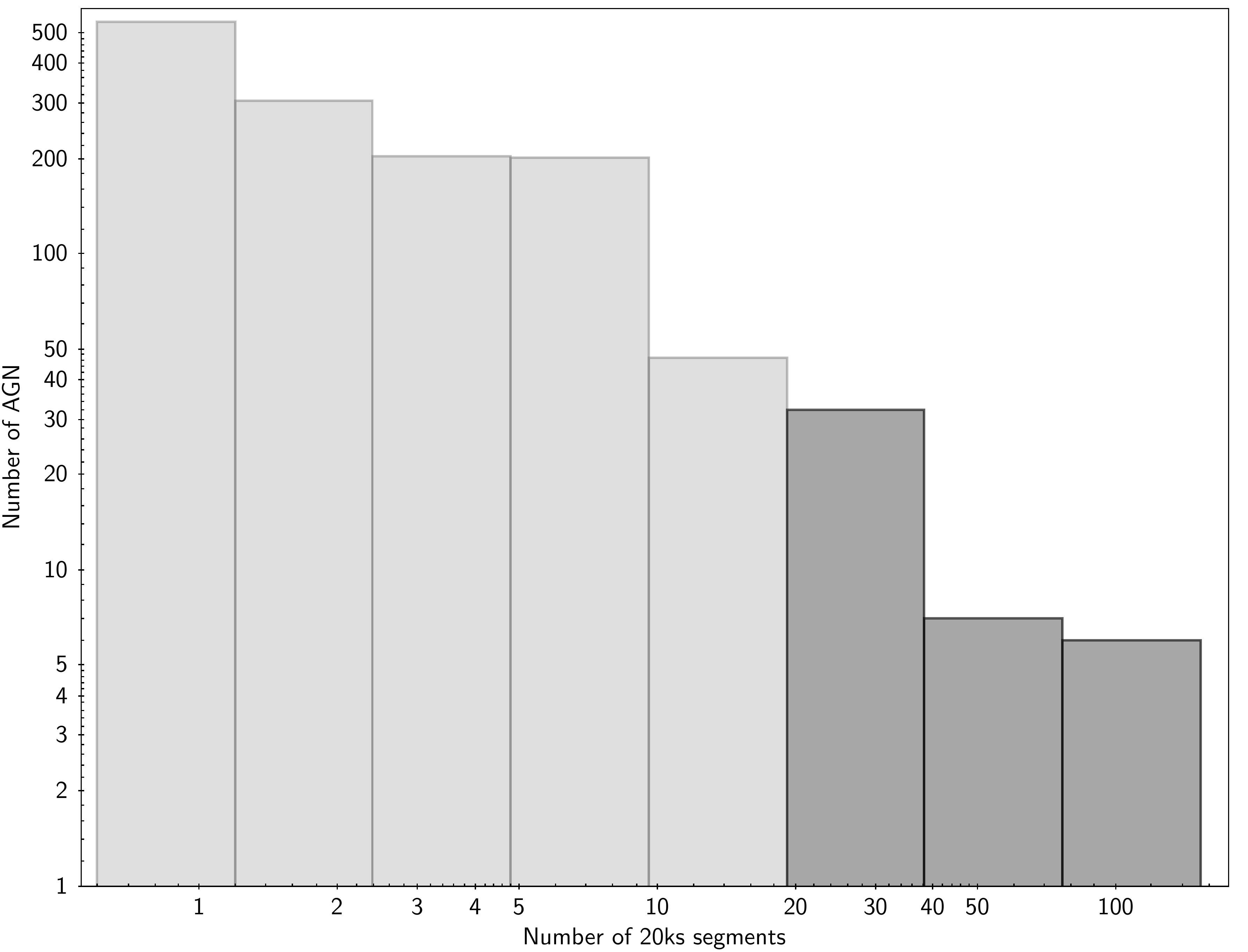}
\caption{Histogram of the number of AGN with \textit{good} 10\,ks and 20\,ks segments from our Sample-S1. The dark bars are for AGN with 20 or more segments from which we create sub-samples S10 and S20 (when $\sigma^2_{\rm{NXS}}$ is positive).}
\label{fig:XCS_NXS_histogram_of_segments}
\end{figure}

Following the method of \citealt{2013ApJ...771....9A} showed, using an in depth statistical analysis of  simulated AGN variability data, that at least 20 {\it good} segments are required for  the $\sigma^2_{\rm{NXS}}$ to be representative of the underlying PSD. We have adopted that constraint in this study. To our knowledge, this is the first time the \citealt{2013ApJ...771....9A} methods have been applied to real variability data.  Figure\,\ref{fig:XCS_NXS_histogram_of_segments} shows number of AGN with \textit{good} 10\,ks, and 20\,ks available from the AGNs in Sample-S1.

\subsection{S10, S20 sub-samples}
From this approach we defined two different sub-samples as S10 and S20. These contain 67 and 45 AGN (18 and 10 with positive $\sigma^2_{\rm{NXS}}$ value) respectively. All $L_{\rm{X}}$, $\sigma^2_{\rm{NXS}}$ (with positive value) and $M_{\rm{BH}}$ from reverberation mapping studies and AGN type are combined and shown in Table\,\ref{tab:A3} in Appendix\,\ref{app:AGN_sampleS10_20_40}. These sub-samples were used to investigate the correlations between $\sigma^2_{\rm{NXS}}$ and $L_{\rm{X}}$ and between $\sigma^2_{\rm{NXS}}$ and $M_{\rm{BH}}$ in Section~\ref{Results}.

With regard to $M_{\rm{BH}}$, these values were taken from \cite{Bentz2015}, using a cross match radius of 5\,arcsec. The number of \cite{Bentz2015} masses for each of the respective sub-samples are listed in Table\,\ref{table:AGNsamples}. With regard to AGN type, these were taken primarily from VC13 (where AGN type is based of the appearance of the Balmer lines) and in the case of two not given in VC13, supplemented by information in the SIMBAD database of astronomical objects \citep{2000A&AS..143....9W}.\\ 



\section{Correlations between X-ray properties and Black Hole mass}
\label{Scaling}
\label{Results}

\begin{figure}
	\includegraphics[width=\columnwidth]{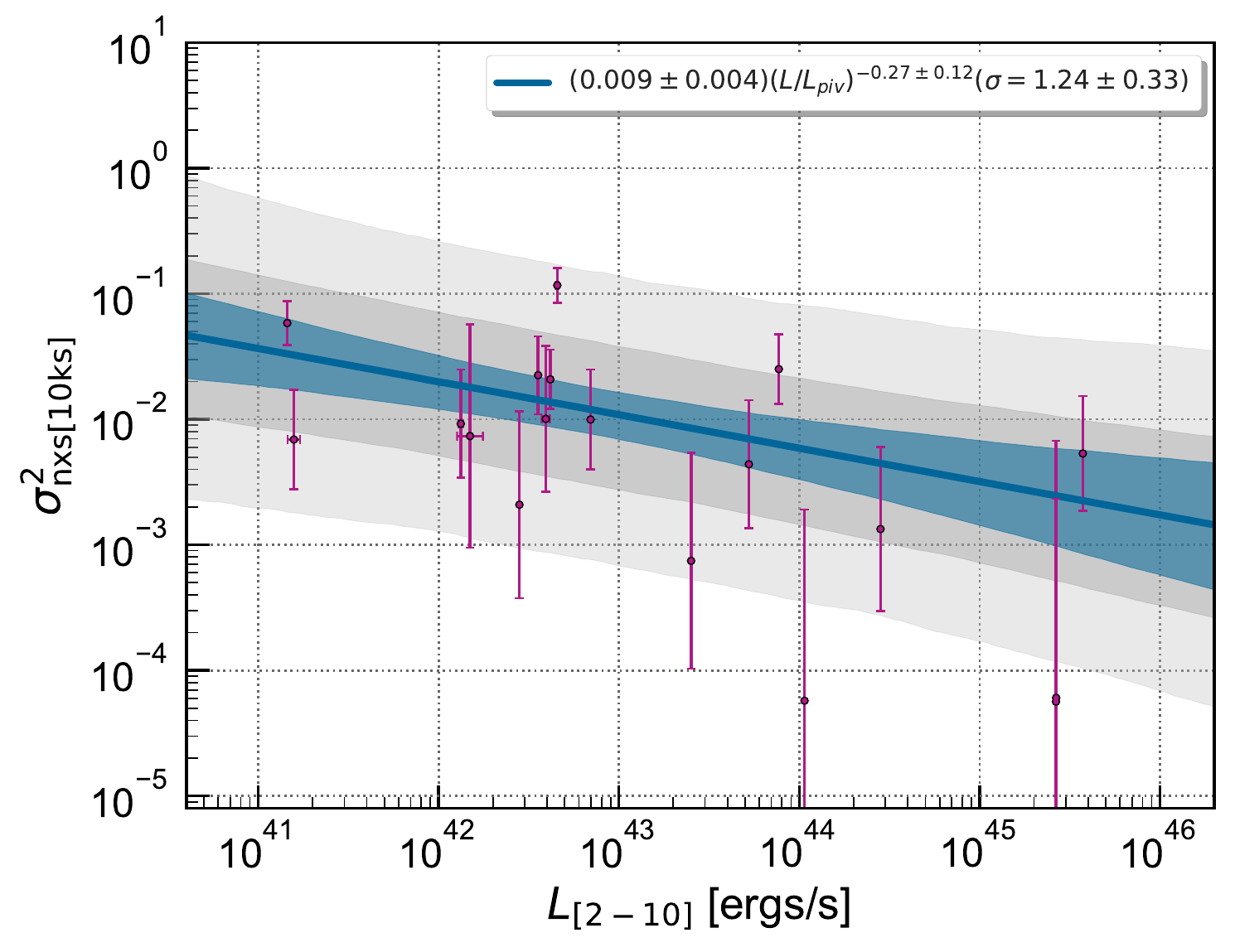}
	\includegraphics[width=\columnwidth]{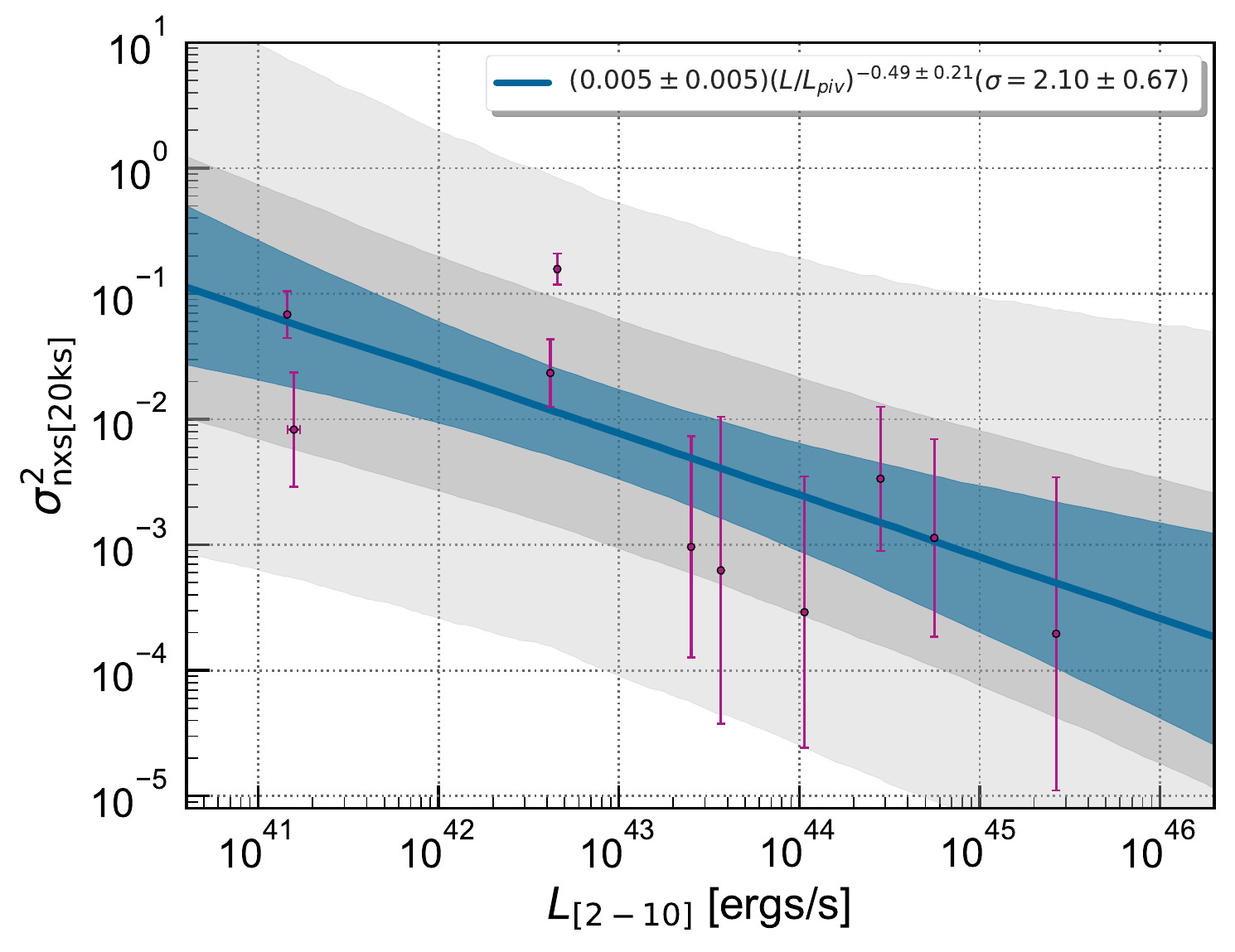}
    \caption{$L_{\rm{X}}$[2-10\,keV] plotted against $\sigma^2_{\rm{NXS}}$ from samples S10 and S20  (i.e. where there are 20 or more 10\,ks and 20\,ks light-curve segments for each AGN). The blue line is the best fit relation with 1-$\sigma$ uncertainty. Grey regions are 1 and 2-$\sigma$ scatter. ${L_{\rm{piv}}=2\times10^{43} \ {\rm erg \ s}^{-1}}$.}
    \label{fig:XCS_hardband_lum_v_NXS_10ks_20ks_40ks_segs}
\end{figure}

\begin{figure}
	\includegraphics[width=\columnwidth]{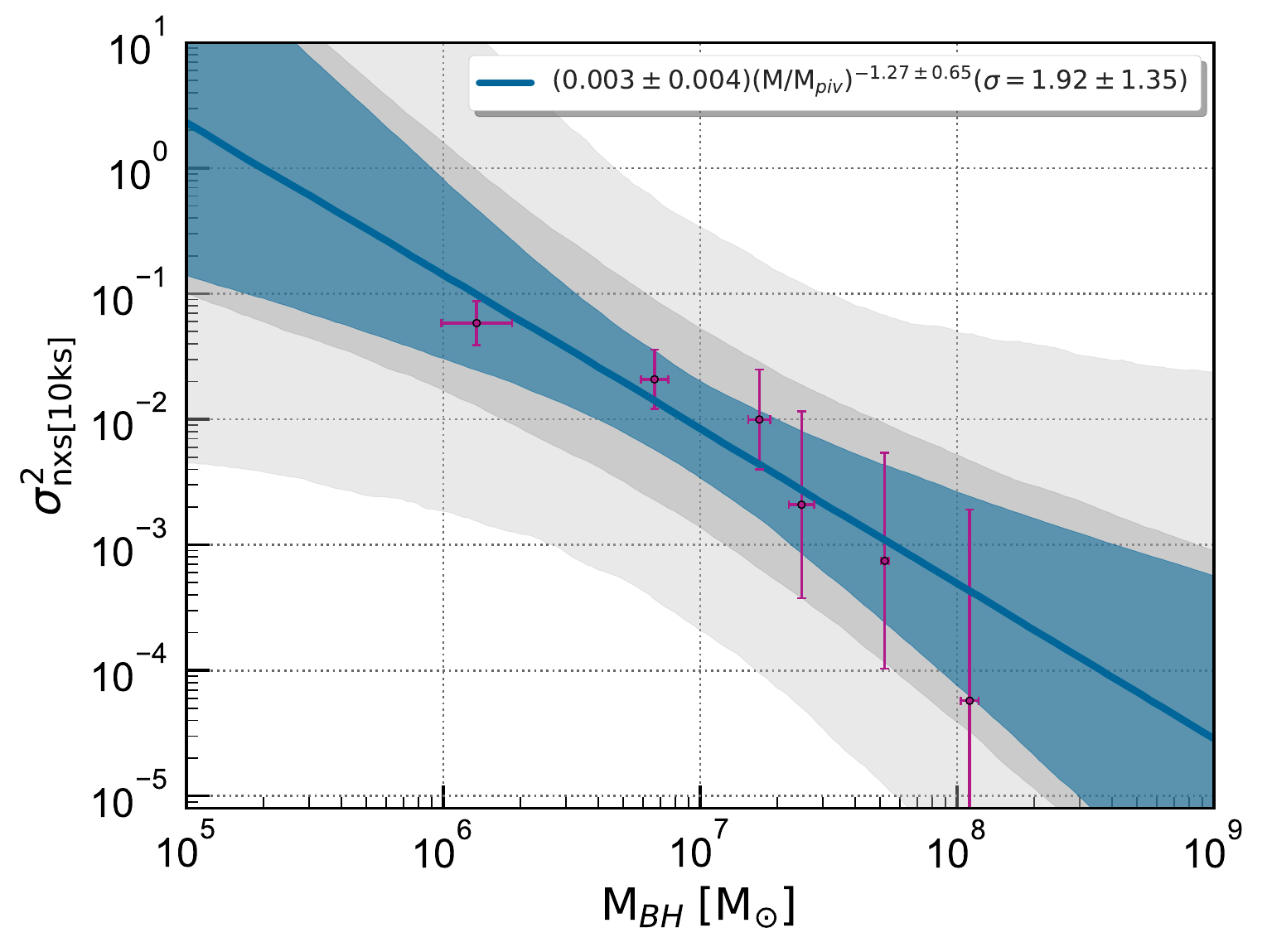}
    \caption{${M}_{\rm{BH}}$ from reverberation mapping studies as catalogued in \citep{Bentz2015} plotted against $\sigma^2_{\rm{NXS}}$  from sample S10 (i.e. where each AGN has 20 or more 10ks light-curve segments). The blue line is the best fit relation with 1-$\sigma$ uncertainty. Grey regions are 1 and $2\sigma$ scatter. ($M_{\rm{piv}}=2\times10^{7}{\rm{M}}_{\odot}$).}
    \label{fig:logM_BH-nxs}
\end{figure}

\begin{figure}
	\includegraphics[width=\columnwidth]{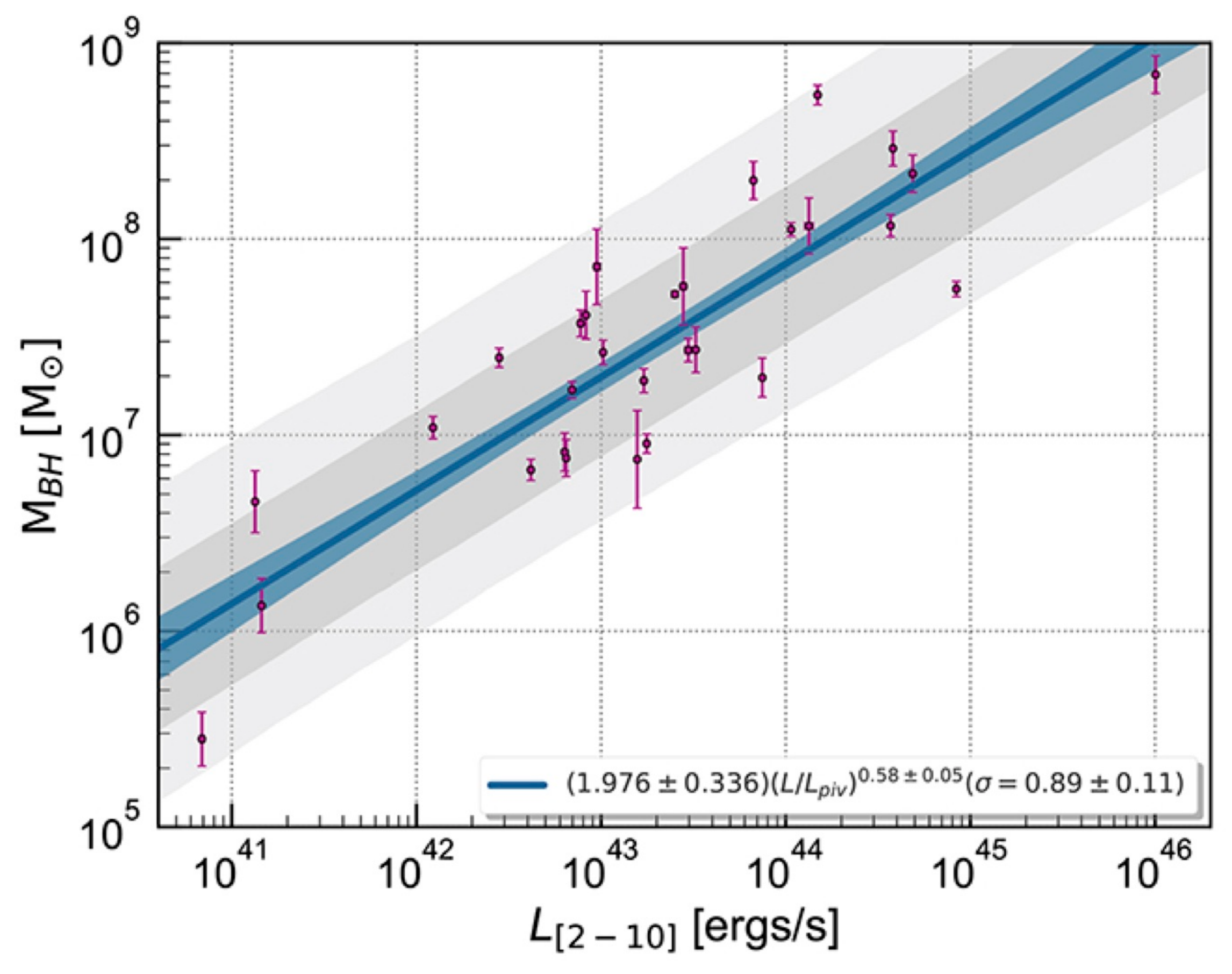}
    \caption{${M}_{\rm{BH}}$ from reverberation mapping studies as catalogued in  \citep{Bentz2015} plotted against $L_{\rm{X}}$[2-10\,keV]. The blue line is the best fit relation with 1-$\sigma$ uncertainty. Grey regions are 1 and 2-$\sigma$ scatter (${L_{\rm{piv}}=2\times10^{43} \ {\rm{erg\,s}^{-1}}}$). }
    \label{fig:XCS_LUM_hard_v_BHM_REVERB}
\end{figure}

We used the regression method of \cite{2007ApJ...665.1489K} to derive the relationships between: {\it i)} $L_{\rm{X}}$ (hard-band 2-10\,keV luminosity) and $\sigma^2_{\rm{NXS}}$ (Figure\,\ref{fig:XCS_hardband_lum_v_NXS_10ks_20ks_40ks_segs} and Table\,\ref{tab:results_LX_nxs}), {\it ii)} $\sigma^2_{\rm{NXS}}$ and $M_{\rm{BH}}$ (Figure\,\ref{fig:logM_BH-nxs}, Table\,\ref{tab:results_BHM_nxs}), and {\it iii)} $L_{\rm{X}}$ and $M_{\rm{BH}}$ (Figure\,\ref{fig:XCS_LUM_hard_v_BHM_REVERB}, Table\,\ref{tab:results_BHM_LX}). 
In order to fit a linear relation to a scaling relation, we adopted the usual practise of fitting to the log of the variables and their measurement errors. 

Significant correlations can be seen in all cases. There is a negative correlation between $\sigma^2_{\rm{NXS}}$ and $L_{\rm{X}}$, i.e. brighter AGN are less variable. There is also a negative correlation between $\sigma^2_{\rm{NXS}}$ and $M_{\rm{BH}}$, i.e. more massive black holes are surrounded by less variable AGN. Not surprisingly, therefore, there is a positive correlation between $L_{\rm{X}}$ and $M_{\rm{BH}}$, i.e. brighter AGN contain more massive black holes. We note that all the $L_{\rm{X}}$ values for the AGN used in these correlations come from on-axis observations (see Section~\ref{sub:lum_estimate}).

\begin{table*}
\begin{center}
\begin{tabular}{c c c c}
 \hline
 Segment duration&Norm.&Slope&Scatter\\[0.5ex]
 [ks]&\textit{A}&$\alpha$&$\sigma$\\
 (1)&(2)&(3)&(4)\\
 \hline
 \hline
 10&0.009$\pm{0.004}$&-0.27$\pm{0.12}$&1.24$\pm{0.33}$\\
 20&0.005$\pm{0.005}$&-0.49$\pm{0.21}$&2.10$\pm{0.67}$\\
 20$^{CAIXA}$&0.007$\pm{0.002}$&-0.60$\pm{0.21}$&1.53$\pm{0.21}$\\
 \hline
 \end{tabular}
  \caption{Correlations between hard-band $L_{\rm{X}}$ and $\sigma^2_{\rm{NXS}}$ of the form log$(\sigma^2_{\rm{NXS}})$ = \textit{A}+$\alpha$log$(L/L_{\rm{piv}})$ where there are 20 or more good light-curve segments for each of the segment durations length in column (1). (2) normalisation, (3) slope, (4) $\sigma$ scatter ${L_{\rm{ piv}}=2\times10^{43}{\rm{erg\,s}^{-1}}}$. \,${CAIXA}$ Results from \citep{2009A&A...495..421B} survey with our fitting methodology.}
 \label{tab:results_LX_nxs}
 \end{center}
\end{table*}

\begin{table*}
\begin{tabular}{c c c}
 \hline
 Normalisation&Slope&Scatter \\[0.5ex]
 \textit{A}&$\alpha$&$\sigma$\\
 (1)&(2)&(3)\\
 \hline
 \hline
 0.003$\pm{0.043}$&-1.27$\pm{0.65}$&1.92$\pm{1.35}$\\
 \hline
 \end{tabular}
 \caption{Correlation between $M_{\rm{BH}}$ from \citep{Bentz2015} and $\sigma^2_{\rm{NXS}}$ of the form log($\sigma^2_{\rm{NXS}})$ = \textit{A}+$\alpha$log$(M/M_{\rm{piv}})$ from sample S10 (i.e where there are 20 or more 10ks light-curve segments for each AGN). (1) normalisation, (2) slope, (3) scatter. $M_{\rm{piv}}=2\times10^{7} M_{\odot}$ }
 \label{tab:results_BHM_nxs}
 \end{table*}

 \begin{table*}
\begin{tabular}{c c c c}
 \hline
 $L_{\rm{X}}$ method&Normalisation&Slope&Scatter\\[0.5ex]
 &\textit{A}&$\alpha$&$\sigma$\\
 (1)&(2)&(3)&(4)\\
 \hline
 \hline
 Spectral fitting from full obs. duration &1.96$\pm{0.336}$&0.58$\pm{0.05}$&0.89$\pm{0.11}$\\
 From count-rate of eight passes of eROSITA duration&1.897$\pm{0.341}$&0.55$\pm{0.05}$&0.90$\pm{0.11}$\\
 \hline
 \end{tabular}
 \caption{Correlations between ${M}_{\rm{BH}}$ from \citep{Bentz2015} and $L_{\rm{X}}$ of the form $\rm{log(M_{BH})}$ = \textit{A}+$\alpha$log$(L/L_{\rm{piv}})$. (1) Method to determine $L_{\rm{X}}$. (2) normalisation, (3) slope, (4) scatter. $L_{\rm{piv}}$=$2\times10^{43}{\rm{erg\,s}^{-1}}$.}
 \label{tab:results_BHM_LX}
 \end{table*}


\begin{figure}
	\includegraphics[width=\columnwidth]{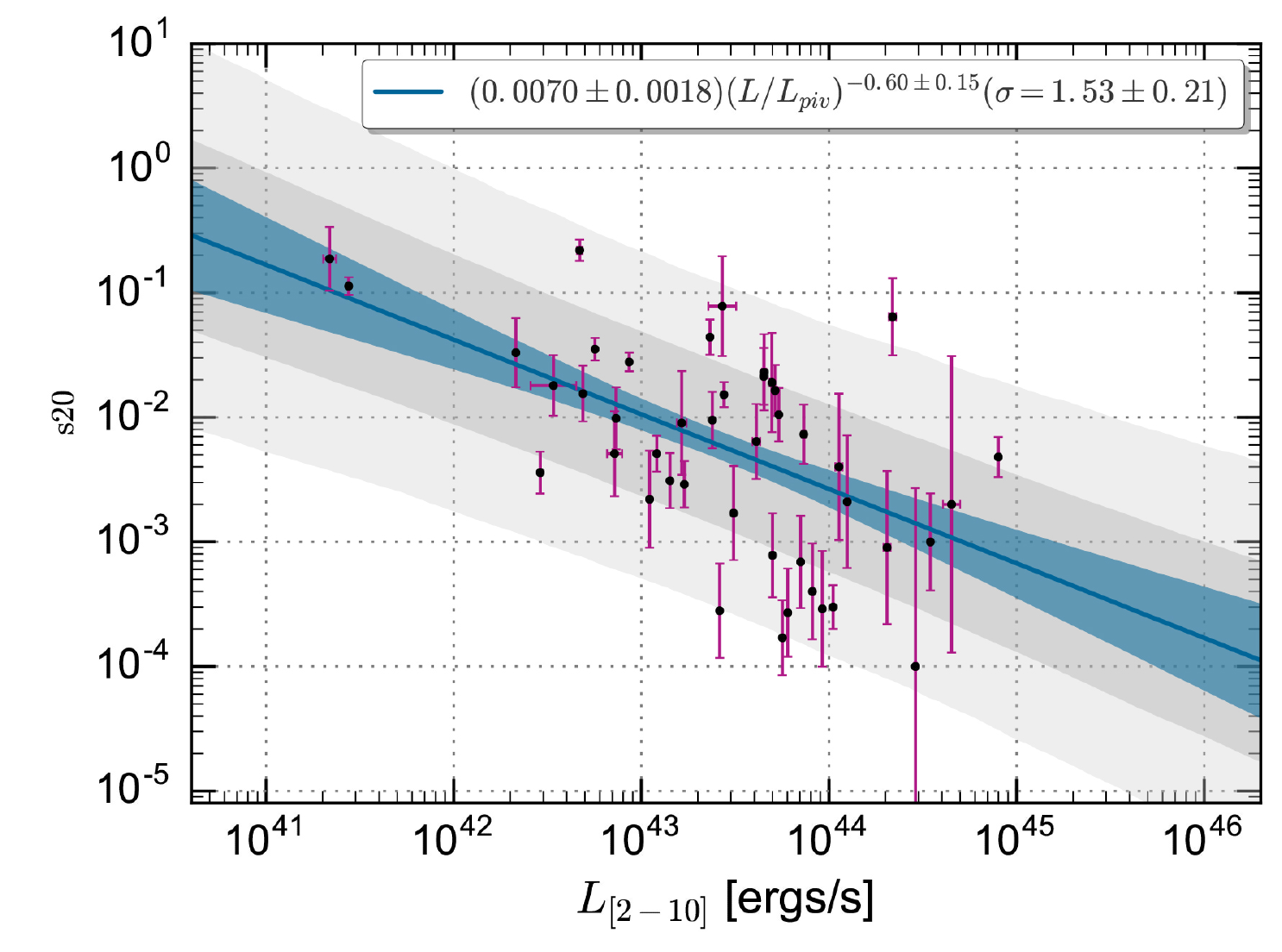}
    \caption{$L_{\rm{X}}$[2-10\,keV] from \citep{2009A&A...501..915B} plotted against $\tt{s20}$ (equivalent to our $\sigma^2_{\rm{NXS20ks}}$ term) for 45 AGN in the CAIXA survey \citep{Ponti2012}. The blue line is the best fit relation with 1-$\sigma$ uncertainty. Grey regions are 1 and 2-$\sigma$ scatter.  ($L_{\rm{piv}}=2\times10^{43}\rm{erg\,s^{-1}}$).}
    \label{fig:CAIXA-logLh-s20}
\end{figure}

Our results are consistent with previous studies that have demonstrated an anti-correlation between luminosity and variability, e.g. \cite{Lawrence1993}, \cite{Barr1986}, \cite{O'Neill2005},  \cite{Ponti2012}. It is not appropriate to compare slopes, scatter and normalisation with published results because of our differing approach to fitting. Instead, we applied our fitting method to the variability and $L_{\rm{X}}$ data in \cite{Ponti2012}, see Figure\,\ref{fig:CAIXA-logLh-s20}. We note that what we call $\sigma^2_{\rm{NXS}}$[20ks], they define as \begin{verb} s20 \end{verb}, albeit with a different approach to error estimation. The fitting results are compared in Table\,\ref{tab:results_LX_nxs} and show good agreement with those shown in the Figure\,\ref{fig:XCS_hardband_lum_v_NXS_10ks_20ks_40ks_segs}.

\section{Implications for eROSITA}
\label{eROSITA}


Due for launch in 2018, the extended Roentgen Survey with an Imaging Telescope Array, eROSITA \citep{2010SPIE.7732E..0UP}, will be the main instrument on the Russian satellite Spektrum-Roentgen-Gamma (SRG). It will be the first X-ray all-sky survey since ROSAT in the 1990s and will observe at X-ray energies between 0.3-10keV. Consisting of seven identical Wolter-1 mirror modules, eROSITA is based on the same pn-CCD technology of XMM, but with smaller pixel size (75$\mu$m, compared to $150\mu$m), better energy resolution ($138{\rm{eV}}$ at $6\rm{keV}$, compared to $\sim150\rm{eV}$ at $6\rm{keV}$) and a larger field of view ($1.03^\circ$ diameter, compared to $0.5^\circ$). eROSITA will carry out an all-sky survey (known as eRASS) over 4-years. The survey will be composed of 8 successive passages over the entire celestial sphere.\\

eRASS is expected to detect up to 3 million AGN out to $z\sim 6$. The potential of these AGN to enhance our understanding, of how galaxies and black holes co-evolve, is enormous.  
However, first, it will be necessary to estimate $M_{\rm{BH}}$ values. As shown in Figure\,\ref{fig:logM_BH-nxs} there is a significant correlation between Type 1 AGNs $L_{\rm{X}}$ and $M_{\rm{BH}}$. In the following we explore whether the $L_{\rm{X}}$ measurements expected from eRASS will be sufficient to be useful in $M_{\rm{BH}}$ estimation.

\subsection{Expectations for eROSITA luminosity measurements}


The eRASS exposure time is dependent on ecliptic latitude ($lat$). According to \cite{2012arXiv1209.3114M}, the approximate exposure time $T_{\rm{EXP}}$ is given by: 
$T_{\rm{EXP}}\sim1627 / \cos(lat)$ seconds for $-84^\circ<lat<84^\circ$ and $T_{\rm {EXP}}$ $\sim17,500$ seconds within $6^\circ$ of the each ecliptic pole. This assumes 100\,per cent observing efficiency. A more realistic efficiency is 80\,per cent. These predictions refer to the full four year survey. The exposure time for each of the eight all-sky surveys, will be 8 times lower, so of the order of hundreds of seconds on average. Therefore, it will be impossible to measure $\sigma^2_{\rm{NXS}}$ values from the majority of eRASS AGN. However, it will still be possible to estimate $L_{\rm{X}}$ values. We forecast the accuracy of the eRASS derived $L_{\rm{X}}$ values below.




\subsubsection{eRASS $L_{\rm{X}}$ from spectral fits}

\label{sec:eRASSLx}

If the AGN flux is sufficiently high, it will be possible to estimate $L_{\rm{X}}$ from the eRASS data using spectral fitting. To predict the accuracy of such fits, we have used the existing XMM observations of AGN in Sample-S1 and selected 2-10\,keV light-curve segments at random, with a duration of the likely eROSITA exposure time in one of the eight All Sky Surveys. The exposure time was adjusted respective to the AGN latitude.  For this exercise we continued to use XMM calibration files, but scaled the exposure time by the ratio of the XMM:eRASS sensitivity \citep[from a comparison of respective effective area in the 2-10\,keV energy range, the combined effective area of the seven eROSITA detectors is a factor of about 3.2 less than the XMM $\tt{PN}$ detector,][]{2012arXiv1209.3114M}.  We extracted source and background spectra for these light-curve segments, and then fit the absorbed powerlaw models as described in Section \ref{sub:lum_estimate}. From these fits we extracted $L_{\rm{X}}$ and $\Delta\LX$ values. 



\begin{figure}
    \includegraphics[width=\columnwidth]{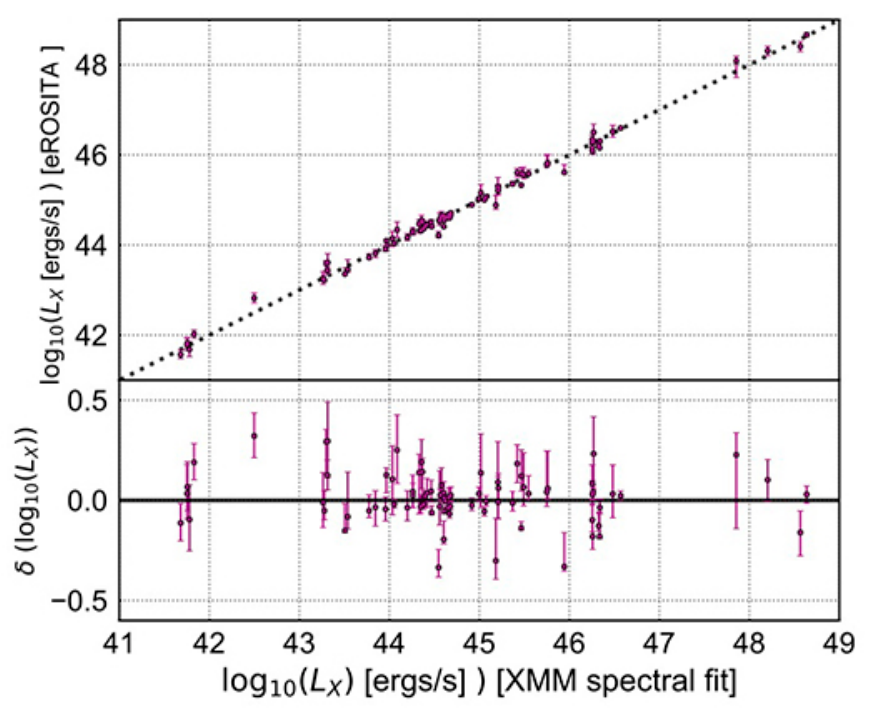}
    \caption{Comparison of $L_{\rm{X}}$ derived from full XMM observation duration spectral fitting of 80 observations (44 AGN) to $L_{\rm{X}}$ derived from estimated eROSITA duration spectral fit. The dotted line shows the one-to-one relation}
    
   \label{fig:erosita_v_XMM_lum}
\end{figure}

Of the AGN in sample S1 tested (1753 XMM observations of 1091 AGN),  successful spectral fits were derived for only 172 observations (corresponding to 98 AGN). Of these, there were 80 observations (44 AGN) where $\frac{\Delta\LX}{\LX}<1$. The $L_{\rm{X}}$ derived from these 80 are compared to those derived from the full XMM exposure time in Figure\,\ref{fig:erosita_v_XMM_lum}. There is excellent agreement albeit only for the $\simeq4\%$ highest flux AGN.

\subsubsection{eRASS $L_{\rm{X}}$ from count-rates}
Where the AGN flux is not high enough to yield a meaningful spectral fit, then it is still possible to estimate $L_{\rm{X}}$ from the source count-rate using an assumed spectral model. For this exercise, we used an absorbed powerlaw (with $\Gamma= 1.7$), with an $n_{\tt{H}}$ value appropriate for the respective AGN galactic latitude. The conversion factors between count-rate and luminosity were generated using XSPEC. For this test we used 254 on-axis observations of 154 AGN. 




To predict an $L_{\rm{X}}$ value for a typical eRASS observation duration, we chose a random start time in the respective observations and set the light-curve duration to be the typical eROSITA observation time at that latitude with a scaling to account for the difference in the XMM:eROSITA sensitivity as used above (Section\,\ref{sec:eRASSLx}). A background subtracted light-curve was extracted in the 2-10\,keV range and the count-rate recorded. We repeated eight times (to mimic the eight eRASS passes) and calculated the mean and error on the mean. These were then converted to $L_{\rm{X}}$ using the scaled XSPEC generated conversion factors. (We note that the error on the mean is likely to be an underestimate since the eight light curves came from the same observation rather than eight different epoch observations as would be the case with eRASS). 

\begin{figure}
    \includegraphics[width=\columnwidth]{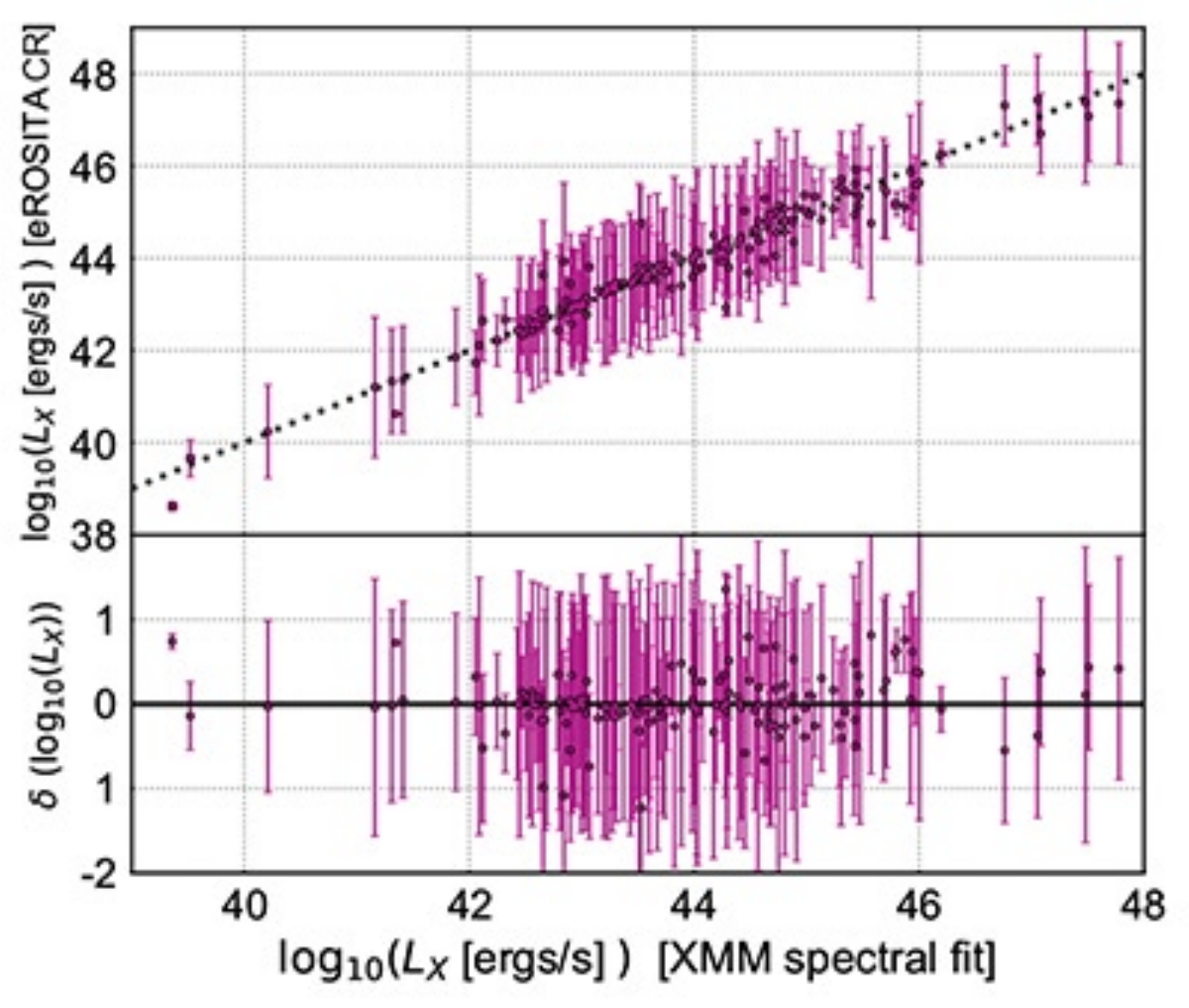}
    \caption{Comparison of $L_{\rm{X}}$  derived from spectral fitting of 154 AGN from XMM full observation time to $L_{\rm{X}}$ derived from fitting using the count-rate method. The dotted line shows the one-to-one relation.}
    
   \label{fig:eRosita_CRlum_vs_XMM_spec}
\end{figure}

Figure\,\ref{fig:eRosita_CRlum_vs_XMM_spec} shows a comparison between $L_{\rm{X}}$ derived from the full-observation spectral method and from this count-rate method (where there were two or more observations of the same AGN, we took the most recent for the count-rate $L_{\rm{X}}$ comparison).

\begin{figure}
	\includegraphics[width=\columnwidth]{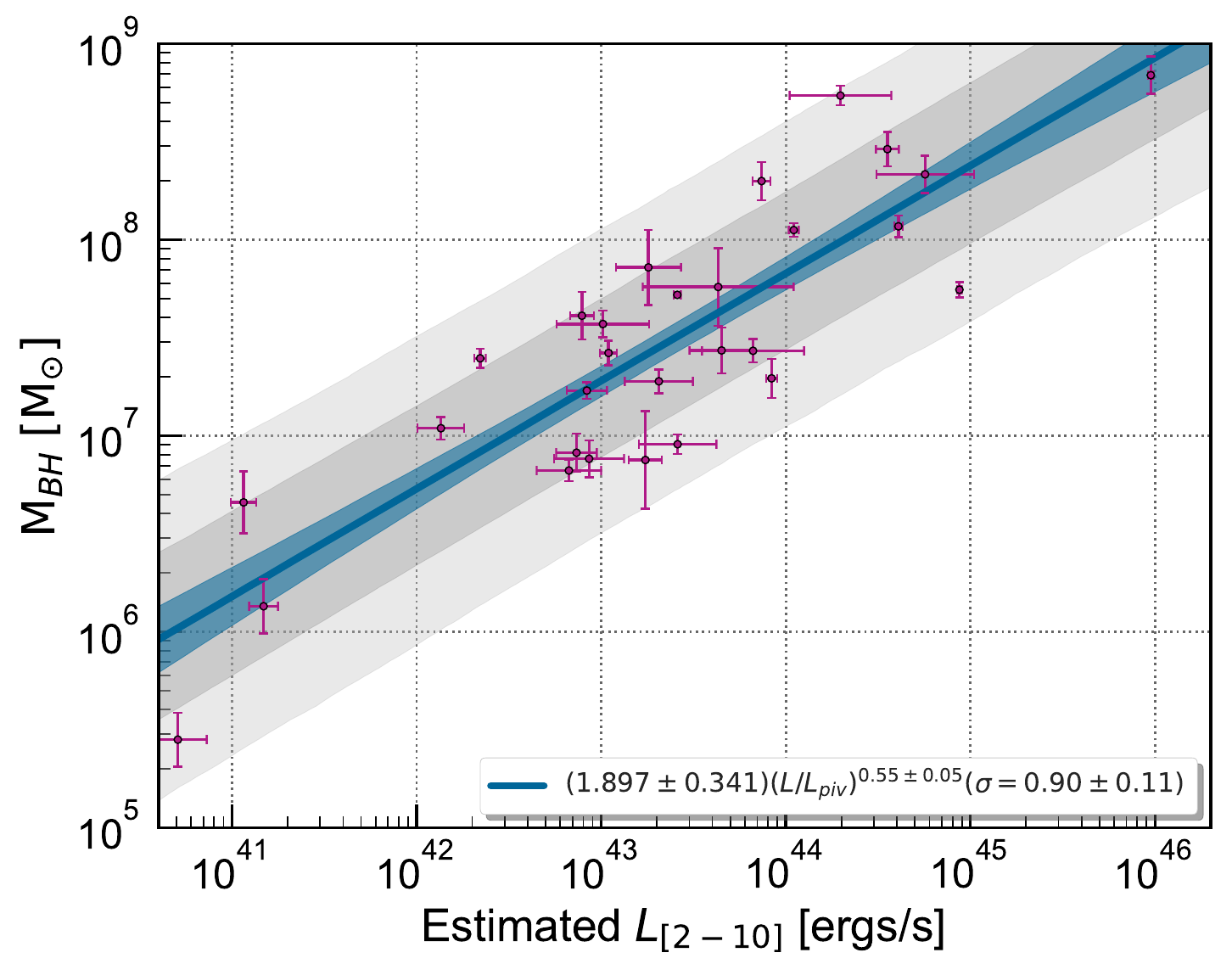}
    \caption{${M}_{\rm{BH}}$ from reverberation mapping studies as catalogued in  \citep{Bentz2015} plotted against hard-band luminosity estimated from the count-rate of eight passes of typical eROSITA exposure duration ($L_{\rm{piv}}=2\times10^{43}\,{\rm{erg\,s^{-1}}}$). The blue line is the best fit relation with 1-$\sigma$ uncertainty. Grey regions are 1 and 2-$\sigma$ scatter. }  
   \label{fig:BHmass_V_erosita_lums}
\end{figure}

We plot the $M_{\rm{BH}}$ to $L_{\rm{X}}$ relation for luminosity derived this way and find that the relation is statistically similar to the relation of $M_{\rm{BH}}$-$L_{\rm{X}}$ from a  spectroscopic analysis of full XMM observations, albeit with larger uncertainty. This is shown in Figure\,\ref{fig:BHmass_V_erosita_lums} with correlation shown in Table\,\ref{tab:results_BHM_LX}. 




\section{Discussion}
\label{Final}
\subsection{AGN Type}
\label{AGN Type}

Our calculations of $L_{\rm{X}}$ assumed that the emission from the AGN is isotropic. This is valid if the emission is not beamed but should be taken into account otherwise. Viewing angle will also have an effect on the line-of-sight hydrogen column density, with Type 2 having higher intrinsic values than Type 1. Therefore, an estimate of the absorption at the location of the AGN itself should be combined with the $n_{\tt{H}}$ value (e.g. from \citealt{Dickey1990}) when the fitting the model.

However, the relationship derived between $L_{\rm{X}}$ and $M_{\rm{BH}}$ shown in Figure\,\ref{fig:XCS_LUM_hard_v_BHM_REVERB} and Table\,\ref{tab:results_BHM_LX} is based only on Type 1 AGN (these are the only ones for which reverberation mapping mass estimates can be made). Therefore, AGN type should not be an issue. That said, when using eRASS $L_{\rm{X}}$ as a proxy for $M_{\rm{BH}}$, one would need to take into account the impact of mixing AGN types. This should not be a problem because spectra will be needed to secure redshifts, and those same spectra can be used to determine AGN type. (A large fraction of the eRASS AGN are planned to be observed by the 4MOST\footnote{{\tt www.4most.eu}} spectrograph).

\subsection{Eddington rate distribution}
The use of $L_{\rm{X}}$ as a proxy for ${M}_{\rm{BH}}$ is dependent on the assumption that AGN radiate at a similar Eddington rate, or at least within a peaked distribution range.
Some authors have shown a peak in the distribution of Eddington ratio. For example, in a study of 407 AGN (in redshift range $\textit{z}\sim0.3 - 4$),  \citealt{2006ApJ...648..128K} showed that the distribution of Eddington ratios is sharply peaked and independent of bolometric luminosity. They also found that at a fixed ${M}_{\rm{BH}}$ the distribution of Eddington ratios is peaked and that this peak occurs at  $L_{\rm{BOL}}/L_{\rm{EDD}}\sim0.25$ with a dispersion of 0.3dex. \citealt{2012MNRAS.425..623L}  found that at high redshift the Eddington ratio distribution is approximately Gaussian,  with a dispersion of $\sim$0.35dex, in Type-1 AGN. \citealt{2010MNRAS.402.2637S} conclude that the distributions of Eddington ratio are similar for AGN (from SDSS) over a wide range of mass and redshift.

\subsection{Selection effect at high redshift}
\label{Selection Effect}
We see from Figure\,\ref{fig:XCS_redshift_v_hLum} that there is a clear trend of increasing $L_{\rm{X}}$ with redshift in sample S1 due to the flux limited nature of the observations. Therefore there may be selection effects that have not so far been taken into account within our correlations involving $L_{\rm{X}}$. Further selection effects may also be involved for relations involving $M_{\rm{BH}}$, as the reverberation-mapped masses are available only for the brightest AGN at relatively low-$z$ ($z<0.23$ in our study).

\subsection{Expanding the sample size}
\label{sect:expanding_sample}
We have based our  $M_{\rm{BH}}$ to $L_{\rm{X}}$ relationship on 30 AGN at $z<0.23$ since these were the only AGN available to us that had $M_{\rm{BH}}$ measurements from reverberation mapping. It would be possible to extend our analysis by including other types of $M_{\rm{BH}}$ measurements e.g. from the luminosity and the width of the broad H$\alpha$ line. For example, there are 220 AGN in our S1 sample where $M_{\rm{BH}}$ has been estimated in \cite{2008Shen} based on H$\beta$, Mg II, and C IV emission lines.

We can also extend our sample by drawing on new compilations of reverberation mapping $M_{\rm{BH}}$ measurements\footnote{We note that late in the preparation of this manuscript, a sample of 44 new $M_{\rm{BH}}$ measurements was published \citep{2017arXiv171103114G}. However, only one of these was new to our S1 sample, and we decided not to repeat the analysis in Section~\ref{Scaling}.}. In particular, we look forward to measurements from the OzDES project \cite{2016arXiv161105456T}. This project is targeting AGN in the Dark Energy Survey deep fields. It aims to derive reverberation mapped $M_{\rm{BH}}$ for $\sim$ 500 AGN over a redshift range of  $0 \ < z < 4$,  with 3\,per cent uncertainty. Cross-matching the OzDES target list with Sample S1, we found 35 AGN in common. Of these 35, fifteen were not already included in Figure\,\ref{fig:XCS_LUM_hard_v_BHM_REVERB}. Of those 15, all but one are at higher redshifts than those included in Figure\,\ref{fig:XCS_LUM_hard_v_BHM_REVERB}. These additional AGN will allow us to test how the correlation between $M_{\rm{BH}}$ and $L_{\rm{X}}$ evolves with cosmic time. Table\,\ref{tab:OzDes_Lum}, we present $L_{\rm{X}}$ values for these 15 derived by following the same spectral fitting methodology described in Section~\ref{Data}.

\begin{table*}
\begin{tabular}{c c c}
 \hline
 XCS Source&z& log($L_{\rm{X}}$) \\ [0.5ex]
 && ${\rm{erg\,s^{-1}}}$ \\
 \hline
  \hline
XMMXCSJ021557.6-045010.3 & 0.884 & 44.03$\pm^{0.12}_{0.02}$\\
XMMXCSJ021628.3-040146.8 & 0.830 & 44.46$\pm^{0.14}_{0.02}$\\
XMMXCSJ021659.7-053204.0 & 2.81 & 45.80$\pm^{0.09}_{0.01}$\\
XMMXCSJ021910.5-055114.0 & 0.558 & 44.22$\pm^{0.09}_{0.01}$\\
XMMXCSJ022024.9-061732.1 & 0.139 & 43.12$\pm^{0.16}_{0.02}$\\
XMMXCSJ022249.5-051453.7 & 0.314 & 44.08$\pm^{0.09}_{0.02}$\\
XMMXCSJ022258.8-045854.8 & 0.466 & 43.69$\pm^{0.30}_{0.04}$\\
XMMXCSJ022415.7-041418.4 & 1.653 & 44.47$\pm^{0.10}_{0.01}$\\
XMMXCSJ022452.1-040519.7 & 0.695 & 43.89$\pm^{0.23}_{0.03}$\\
XMMXCSJ022711.8-045037.4 & 0.961 & 44.49$\pm^{0.06}_{0.02}$\\
XMMXCSJ022716.1-044537.6 & 0.721 & 44.53$\pm^{0.10}_{0.02}$\\
XMMXCSJ022845.6-043350.7 & 1.865 & 45.50$\pm^{0.07}_{0.01}$\\
XMMXCSJ022851.4-051224.4 & 0.316 & 44.13$\pm^{0.05}_{0.03}$\\
XMMXCSJ033208.9-274732.1 & 0.544 & 43.91$\pm^{0.03}_{0.02}$\\
XMMXCSJ033211.5-273727.8 & 1.570 & 44.48$\pm^{0.25}_{0.00}$\\
 \hline
\end{tabular}
\caption{Estimated hard-band $L_{\rm{X}}$, from spectral fitting, of AGN in the OzDES reverberation mapping survey. These objects did not form part of the samples described in Table\,\ref{table:AGNsamples}.}
\label{tab:OzDes_Lum}
\end{table*}

\section{Summary}

In this paper we used AGN associated with XCS point sources to confirm the existence of scaling relations between $M_{\rm{BH}}$ and $\sigma^2_{\rm{NXS}}$, and between $L_{\rm{X}}$ and $\sigma^2_{\rm{NXS}}$. We have also demonstrated preliminary evidence for a correlation between  $M_{\rm{BH}}$ and $L_{\rm{X}}$. Such a correlation, if confirmed, would open up the possibility of estimating distributions of $M_{\rm{BH}}$  values for 100's of thousands of AGN detected during the eRASS survey by eROSITA. We note that the scatter in the relation is large: at a given $L_{\rm{X}}$, the $M_{\rm{BH}}$ can vary by up to two orders of magnitude. Therefore, the method would not be suitable to measure $M_{\rm{BH}}$ for individual objects, but rather of ensemble populations.\\

We have described a method to estimate the $L_{\rm{X}}$ of an AGN from count-rates of short duration observations, such as those in the eRASS, where $\sigma^2_{\rm{NXS}}$ cannot be measured. We have shown that although the uncertainties on the count-rate derived $L_{\rm{X}}$ are larger than spectrally derived $L_{\rm{X}}$, the scaling relation with $M_{\rm{BH}}$ is the statistically similar. \\

We have estimated that the number of reverbation mapping derived $M_{\rm{BH}}$ estimates for AGN with XMM-Newton derived $L_{\rm{X}}$ values will soon increase by up to 50\% thanks to the OzDES project. Almost all of these new $M_{\rm{BH}}$ values will be for AGN beyond the redshift grasp of our current sample. Testing whether the $M_{\rm{BH}}$ to $L_{\rm{X}}$ correlation persists at other epochs will be essential before this method can be applied with confidence to the eRASS AGN sample.\\

    \section*{Acknowledgements}
    
    JM acknowledges support from MPS, University of Sussex. 
    
    KR acknowledges support from the Science and Technology Facilities Council (grant number ST/P000252/1).

    AF acknowledges support from the McWilliams Postdoctoral Fellowship.

    MS acknowledges support by the Olle Engkvist Foundation (Stiftelsen Olle Engkvist Byggm\"astare).
    
    MH acknowledges financial support from the National Research Foundation, the South African Square Kilometre Array project, and the University of KwaZulu-Natal.

\bibliographystyle{mnras}
\bibliography{bib} 

\begin{thebibliography}{}
\makeatletter
\relax
\def\mn@urlcharsother{\let\do\@makeother \do\$\do\&\do\#\do\^\do\_\do\%\do\~}
\def\mn@doi{\begingroup\mn@urlcharsother \@ifnextchar [ {\mn@doi@}
  {\mn@doi@[]}}
\def\mn@doi@[#1]#2{\def\@tempa{#1}\ifx\@tempa\@empty \href
  {http://dx.doi.org/#2} {doi:#2}\else \href {http://dx.doi.org/#2} {#1}\fi
  \endgroup}
\def\mn@eprint#1#2{\mn@eprint@#1:#2::\@nil}
\def\mn@eprint@arXiv#1{\href {http://arxiv.org/abs/#1} {{\tt arXiv:#1}}}
\def\mn@eprint@dblp#1{\href {http://dblp.uni-trier.de/rec/bibtex/#1.xml}
  {dblp:#1}}
\def\mn@eprint@#1:#2:#3:#4\@nil{\def\@tempa {#1}\def\@tempb {#2}\def\@tempc
  {#3}\ifx \@tempc \@empty \let \@tempc \@tempb \let \@tempb \@tempa \fi \ifx
  \@tempb \@empty \def\@tempb {arXiv}\fi \@ifundefined
  {mn@eprint@\@tempb}{\@tempb:\@tempc}{\expandafter \expandafter \csname
  mn@eprint@\@tempb\endcsname \expandafter{\@tempc}}}

\bibitem[\protect\citeauthoryear{{Allevato}, {Paolillo}, {Papadakis}  \&
  {Pinto}}{{Allevato} et~al.}{2013}]{2013ApJ...771....9A}
{Allevato} V.,  {Paolillo} M.,  {Papadakis} I.,   {Pinto} C.,  2013, \mn@doi
  [\apj] {10.1088/0004-637X/771/1/9}, \href
  {http://adsabs.harvard.edu/abs/2013ApJ...771....9A} {771, 9}

\bibitem[\protect\citeauthoryear{{Antonucci}, {Vagnetti}  \&
  {Trevese}}{{Antonucci} et~al.}{2014}]{Antonucci2014}
{Antonucci} M.,  {Vagnetti} F.,   {Trevese} D.,  2014, in The X-ray Universe
  2014. p.~222

\bibitem[\protect\citeauthoryear{{Barr} \& {Mushotzky}}{{Barr} \&
  {Mushotzky}}{1986}]{Barr1986}
{Barr} P.,  {Mushotzky} R.~F.,  1986, \mn@doi [\nat] {10.1038/320421a0}, \href
  {http://adsabs.harvard.edu/abs/1986Natur.320..421B} {320, 421}

\bibitem[\protect\citeauthoryear{{Bentz} \& {Katz}}{{Bentz} \&
  {Katz}}{2015}]{Bentz2015}
{Bentz} M.~C.,  {Katz} S.,  2015, \mn@doi [\pasp] {10.1086/679601}, \href
  {http://adsabs.harvard.edu/abs/2015PASP..127...67B} {127, 67}

\bibitem[\protect\citeauthoryear{{Bianchi}, {Guainazzi}, {Matt}, {Fonseca
  Bonilla}  \& {Ponti}}{{Bianchi} et~al.}{2009a}]{2009A&A...495..421B}
{Bianchi} S.,  {Guainazzi} M.,  {Matt} G.,  {Fonseca Bonilla} N.,   {Ponti} G.,
   2009a, \mn@doi [\aap] {10.1051/0004-6361:200810620}, \href
  {http://adsabs.harvard.edu/abs/2009A%26A...495..421B} {495, 421}

\bibitem[\protect\citeauthoryear{{Bianchi}, {Bonilla}, {Guainazzi}, {Matt}  \&
  {Ponti}}{{Bianchi} et~al.}{2009b}]{2009A&A...501..915B}
{Bianchi} S.,  {Bonilla} N.~F.,  {Guainazzi} M.,  {Matt} G.,   {Ponti} G.,
  2009b, \mn@doi [\aap] {10.1051/0004-6361/200911905}, \href
  {http://adsabs.harvard.edu/abs/2009A%26A...501..915B} {501, 915}

\bibitem[\protect\citeauthoryear{{Blandford} \& {McKee}}{{Blandford} \&
  {McKee}}{1982}]{Blandford1982}
{Blandford} R.~D.,  {McKee} C.~F.,  1982, \mn@doi [\apj] {10.1086/159843},
  \href {http://adsabs.harvard.edu/abs/1982ApJ...255..419B} {255, 419}

\bibitem[\protect\citeauthoryear{{Corral}, {Della Ceca}, {Caccianiga},
  {Severgnini}, {Brunner}, {Carrera}, {Page}  \& {Schwope}}{{Corral}
  et~al.}{2011}]{2011A&A...530A..42C}
{Corral} A.,  {Della Ceca} R.,  {Caccianiga} A.,  {Severgnini} P.,  {Brunner}
  H.,  {Carrera} F.~J.,  {Page} M.~J.,   {Schwope} A.~D.,  2011, \mn@doi [\aap]
  {10.1051/0004-6361/201015227}, \href
  {http://adsabs.harvard.edu/abs/2011A%26A...530A..42C} {530, A42}

\bibitem[\protect\citeauthoryear{{DeGraf}, {Di Matteo}, {Treu}, {Feng}, {Woo}
  \& {Park}}{{DeGraf} et~al.}{2015}]{2015MNRAS.454..913D}
{DeGraf} C.,  {Di Matteo} T.,  {Treu} T.,  {Feng} Y.,  {Woo} J.-H.,   {Park}
  D.,  2015, \mn@doi [\mnras] {10.1093/mnras/stv2002}, \href
  {http://adsabs.harvard.edu/abs/2015MNRAS.454..913D} {454, 913}

\bibitem[\protect\citeauthoryear{{Dickey} \& {Lockman}}{{Dickey} \&
  {Lockman}}{1990}]{Dickey1990}
{Dickey} J.~M.,  {Lockman} F.~J.,  1990, \mn@doi [\araa]
  {10.1146/annurev.aa.28.090190.001243}, \href
  {http://adsabs.harvard.edu/abs/1990ARA%26A..28..215D} {28, 215}

\bibitem[\protect\citeauthoryear{{Elvis}, {Maccacaro}, {Wilson}, {Ward},
  {Penston}, {Fosbury}  \& {Perola}}{{Elvis} et~al.}{1978}]{Elvis1978}
{Elvis} M.,  {Maccacaro} T.,  {Wilson} A.~S.,  {Ward} M.~J.,  {Penston} M.~V.,
  {Fosbury} R.~A.~E.,   {Perola} G.~C.,  1978, \mn@doi [\mnras]
  {10.1093/mnras/183.2.129}, \href
  {http://adsabs.harvard.edu/abs/1978MNRAS.183..129E} {183, 129}

\bibitem[\protect\citeauthoryear{{Ferrarese} \& {Ford}}{{Ferrarese} \&
  {Ford}}{2005}]{2005SSRv..116..523F}
{Ferrarese} L.,  {Ford} H.,  2005, \mn@doi [\ssr] {10.1007/s11214-005-3947-6},
  \href {http://adsabs.harvard.edu/abs/2005SSRv..116..523F} {116, 523}

\bibitem[\protect\citeauthoryear{{Gandhi}}{{Gandhi}}{2005}]{Gandhi2005}
{Gandhi} P.,  2005, Asian Journal of Physics, \href
  {http://adsabs.harvard.edu/abs/2005AsJPh..13...90G} {13, 90}

\bibitem[\protect\citeauthoryear{{Gebhardt} et~al.,}{{Gebhardt}
  et~al.}{2000}]{2000ApJ...539L..13G}
{Gebhardt} K.,  et~al., 2000, \mn@doi [\apjl] {10.1086/312840}, \href
  {http://adsabs.harvard.edu/abs/2000ApJ...539L..13G} {539, L13}

\bibitem[\protect\citeauthoryear{{Ghez}, {Salim}, {Hornstein}, {Tanner}, {Lu},
  {Morris}, {Becklin}  \& {Duch{\^e}ne}}{{Ghez}
  et~al.}{2005}]{2005ApJ...620..744G}
{Ghez} A.~M.,  {Salim} S.,  {Hornstein} S.~D.,  {Tanner} A.,  {Lu} J.~R.,
  {Morris} M.,  {Becklin} E.~E.,   {Duch{\^e}ne} G.,  2005, \mn@doi [\apj]
  {10.1086/427175}, \href {http://adsabs.harvard.edu/abs/2005ApJ...620..744G}
  {620, 744}

\bibitem[\protect\citeauthoryear{{Grier} et~al.,}{{Grier}
  et~al.}{2012}]{2012ApJ...755...60G}
{Grier} C.~J.,  et~al., 2012, \mn@doi [\apj] {10.1088/0004-637X/755/1/60},
  \href {http://adsabs.harvard.edu/abs/2012ApJ...755...60G} {755, 60}

\bibitem[\protect\citeauthoryear{{Grier} et~al.,}{{Grier}
  et~al.}{2017}]{2017arXiv171103114G}
{Grier} C.~J.,  et~al., 2017, preprint, \href
  {http://adsabs.harvard.edu/abs/2017arXiv171103114G} {} (\mn@eprint {arXiv}
  {1711.03114})

\bibitem[\protect\citeauthoryear{{Kamizasa}, {Terashima}  \&
  {Awaki}}{{Kamizasa} et~al.}{2012}]{Kamizasa2012}
{Kamizasa} N.,  {Terashima} Y.,   {Awaki} H.,  2012, \mn@doi [\apj]
  {10.1088/0004-637X/751/1/39}, \href
  {http://adsabs.harvard.edu/abs/2012ApJ...751...39K} {751, 39}

\bibitem[\protect\citeauthoryear{{Kaspi}, {Smith}, {Netzer}, {Maoz}, {Jannuzi}
  \& {Giveon}}{{Kaspi} et~al.}{2000}]{2000ApJ...533..631K}
{Kaspi} S.,  {Smith} P.~S.,  {Netzer} H.,  {Maoz} D.,  {Jannuzi} B.~T.,
  {Giveon} U.,  2000, \mn@doi [\apj] {10.1086/308704}, \href
  {http://adsabs.harvard.edu/abs/2000ApJ...533..631K} {533, 631}

\bibitem[\protect\citeauthoryear{{Kelly}}{{Kelly}}{2007}]{2007ApJ...665.1489K}
{Kelly} B.~C.,  2007, \mn@doi [\apj] {10.1086/519947}, \href
  {http://adsabs.harvard.edu/abs/2007ApJ...665.1489K} {665, 1489}

\bibitem[\protect\citeauthoryear{{Kelly}, {Treu}, {Malkan}, {Pancoast}  \&
  {Woo}}{{Kelly} et~al.}{2013}]{2013ApJ...779..187K}
{Kelly} B.~C.,  {Treu} T.,  {Malkan} M.,  {Pancoast} A.,   {Woo} J.-H.,  2013,
  \mn@doi [\apj] {10.1088/0004-637X/779/2/187}, \href
  {http://adsabs.harvard.edu/abs/2013ApJ...779..187K} {779, 187}

\bibitem[\protect\citeauthoryear{{Kollmeier} et~al.,}{{Kollmeier}
  et~al.}{2006}]{2006ApJ...648..128K}
{Kollmeier} J.~A.,  et~al., 2006, \mn@doi [\apj] {10.1086/505646}, \href
  {http://adsabs.harvard.edu/abs/2006ApJ...648..128K} {648, 128}

\bibitem[\protect\citeauthoryear{{K{\"o}rding}, {Migliari}, {Fender},
  {Belloni}, {Knigge}  \& {McHardy}}{{K{\"o}rding}
  et~al.}{2007}]{2007MNRAS.380..301K}
{K{\"o}rding} E.~G.,  {Migliari} S.,  {Fender} R.,  {Belloni} T.,  {Knigge} C.,
    {McHardy} I.,  2007, \mn@doi [\mnras] {10.1111/j.1365-2966.2007.12067.x},
  \href {http://adsabs.harvard.edu/abs/2007MNRAS.380..301K} {380, 301}

\bibitem[\protect\citeauthoryear{{Koz{\l}owski}}{{Koz{\l}owski}}{2016}]{2016arXiv160909489K}
{Koz{\l}owski} S.,  2016, preprint, \href
  {http://adsabs.harvard.edu/abs/2016arXiv160909489K} {} (\mn@eprint {arXiv}
  {1609.09489})

\bibitem[\protect\citeauthoryear{{Lawrence} \& {Papadakis}}{{Lawrence} \&
  {Papadakis}}{1993}]{Lawrence1993}
{Lawrence} A.,  {Papadakis} I.,  1993, \mn@doi [\apjl] {10.1086/187002}, \href
  {http://adsabs.harvard.edu/abs/1993ApJ...414L..85L} {414, L85}

\bibitem[\protect\citeauthoryear{{Lloyd-Davies} et~al.,}{{Lloyd-Davies}
  et~al.}{2011}]{LloydDavies2011}
{Lloyd-Davies} E.~J.,  et~al., 2011, \mn@doi [mnras]
  {10.1111/j.1365-2966.2011.19117.x}, \href
  {http://adsabs.harvard.edu/abs/2011MNRAS.418...14L} {418, 14}

\bibitem[\protect\citeauthoryear{{Ludlam}, {Cackett}, {G{\"u}ltekin}, {Fabian},
  {Gallo}  \& {Miniutti}}{{Ludlam} et~al.}{2015}]{2015MNRAS.447.2112L}
{Ludlam} R.~M.,  {Cackett} E.~M.,  {G{\"u}ltekin} K.,  {Fabian} A.~C.,  {Gallo}
  L.,   {Miniutti} G.,  2015, \mn@doi [\mnras] {10.1093/mnras/stu2618}, \href
  {http://adsabs.harvard.edu/abs/2015MNRAS.447.2112L} {447, 2112}

\bibitem[\protect\citeauthoryear{{Lusso} et~al.,}{{Lusso}
  et~al.}{2012}]{2012MNRAS.425..623L}
{Lusso} E.,  et~al., 2012, \mn@doi [\mnras] {10.1111/j.1365-2966.2012.21513.x},
  \href {http://adsabs.harvard.edu/abs/2012MNRAS.425..623L} {425, 623}

\bibitem[\protect\citeauthoryear{{Markowitz} et~al.,}{{Markowitz}
  et~al.}{2003}]{2003ApJ...593...96M}
{Markowitz} A.,  et~al., 2003, \mn@doi [\apj] {10.1086/375330}, \href
  {http://adsabs.harvard.edu/abs/2003ApJ...593...96M} {593, 96}

\bibitem[\protect\citeauthoryear{{McHardy}}{{McHardy}}{1988}]{1988MmSAI..59..239M}
{McHardy} I.,  1988, \memsai, \href
  {http://adsabs.harvard.edu/abs/1988MmSAI..59..239M} {59, 239}

\bibitem[\protect\citeauthoryear{{McHardy}, {Koerding}, {Knigge}, {Uttley}  \&
  {Fender}}{{McHardy} et~al.}{2006}]{2006Natur.444..730M}
{McHardy} I.~M.,  {Koerding} E.,  {Knigge} C.,  {Uttley} P.,   {Fender} R.~P.,
  2006, \mn@doi [\nat] {10.1038/nature05389}, \href
  {http://adsabs.harvard.edu/abs/2006Natur.444..730M} {444, 730}

\bibitem[\protect\citeauthoryear{{Merloni} et~al.,}{{Merloni}
  et~al.}{2012}]{2012arXiv1209.3114M}
{Merloni} A.,  et~al., 2012, preprint, \href
  {http://adsabs.harvard.edu/abs/2012arXiv1209.3114M} {} (\mn@eprint {arXiv}
  {1209.3114})

\bibitem[\protect\citeauthoryear{{Middei}, {Vagnetti}, {Antonucci}  \&
  {Serafinelli}}{{Middei} et~al.}{2016}]{2016JPhCS.689a2006M}
{Middei} R.,  {Vagnetti} F.,  {Antonucci} M.,   {Serafinelli} R.,  2016,
  \mn@doi [Journal of Physics Conference Series]
  {10.1088/1742-6596/689/1/012006}, \href
  {http://adsabs.harvard.edu/abs/2016JPhCS.689a2006M} {689, 012006}

\bibitem[\protect\citeauthoryear{{Nandra} \& {Pounds}}{{Nandra} \&
  {Pounds}}{1994}]{Nandra1994}
{Nandra} K.,  {Pounds} K.~A.,  1994, \mn@doi [\mnras]
  {10.1093/mnras/268.2.405}, \href
  {http://adsabs.harvard.edu/abs/1994MNRAS.268..405N} {268, 405}

\bibitem[\protect\citeauthoryear{{Nandra}, {George}, {Mushotzky}, {Turner}  \&
  {Yaqoob}}{{Nandra} et~al.}{1997}]{Nandra1997}
{Nandra} K.,  {George} I.~M.,  {Mushotzky} R.~F.,  {Turner} T.~J.,   {Yaqoob}
  T.,  1997, \apj, \href {http://adsabs.harvard.edu/abs/1997ApJ...476...70N}
  {476, 70}

\bibitem[\protect\citeauthoryear{{Nikolajuk}, {Papadakis}  \&
  {Czerny}}{{Nikolajuk} et~al.}{2004}]{2004MNRAS.350L..26N}
{Nikolajuk} M.,  {Papadakis} I.~E.,   {Czerny} B.,  2004, \mn@doi [\mnras]
  {10.1111/j.1365-2966.2004.07829.x}, \href
  {http://adsabs.harvard.edu/abs/2004MNRAS.350L..26N} {350, L26}

\bibitem[\protect\citeauthoryear{{O'Neill}, {Nandra}, {Papadakis}  \&
  {Turner}}{{O'Neill} et~al.}{2005}]{O'Neill2005}
{O'Neill} P.~M.,  {Nandra} K.,  {Papadakis} I.~E.,   {Turner} T.~J.,  2005,
  \mn@doi [\mnras] {10.1111/j.1365-2966.2005.08860.x}, \href
  {http://adsabs.harvard.edu/abs/2005MNRAS.358.1405O} {358, 1405}

\bibitem[\protect\citeauthoryear{{Pan}, {Yuan}, {Zhou}, {Dong}  \& {Liu}}{{Pan}
  et~al.}{2015}]{2015ApJ...808..163P}
{Pan} H.-W.,  {Yuan} W.,  {Zhou} X.-L.,  {Dong} X.-B.,   {Liu} B.,  2015,
  \mn@doi [\apj] {10.1088/0004-637X/808/2/163}, \href
  {http://adsabs.harvard.edu/abs/2015ApJ...808..163P} {808, 163}

\bibitem[\protect\citeauthoryear{{Papadakis}}{{Papadakis}}{2004}]{Papadakis2004}
{Papadakis} I.~E.,  2004, \mn@doi [\mnras] {10.1111/j.1365-2966.2004.07351.x},
  \href {http://adsabs.harvard.edu/abs/2004MNRAS.348..207P} {348, 207}

\bibitem[\protect\citeauthoryear{{Peterson}, {Wanders}, {Bertram}, {Hunley},
  {Pogge}  \& {Wagner}}{{Peterson} et~al.}{1998}]{1998ApJ...501...82P}
{Peterson} B.~M.,  {Wanders} I.,  {Bertram} R.,  {Hunley} J.~F.,  {Pogge}
  R.~W.,   {Wagner} R.~M.,  1998, \mn@doi [\apj] {10.1086/305813}, \href
  {http://adsabs.harvard.edu/abs/1998ApJ...501...82P} {501, 82}

\bibitem[\protect\citeauthoryear{{Ponti}, {Papadakis}, {Bianchi}, {Guainazzi},
  {Matt}, {Uttley}  \& {Bonilla}}{{Ponti} et~al.}{2012}]{Ponti2012}
{Ponti} G.,  {Papadakis} I.,  {Bianchi} S.,  {Guainazzi} M.,  {Matt} G.,
  {Uttley} P.,   {Bonilla} N.~F.,  2012, \mn@doi [\aap]
  {10.1051/0004-6361/201118326}, \href
  {http://adsabs.harvard.edu/abs/2012A%26A...542A..83P} {542, A83}

\bibitem[\protect\citeauthoryear{{Pounds} \& {McHardy}}{{Pounds} \&
  {McHardy}}{1988}]{1988pnsb.conf..285P}
{Pounds} K.~A.,  {McHardy} I.~M.,  1988, in {Tanaka} Y.,  ed., Physics of
  Neutron Stars and Black Holes. pp 285--299

\bibitem[\protect\citeauthoryear{{Predehl} et~al.,}{{Predehl}
  et~al.}{2010}]{2010SPIE.7732E..0UP}
{Predehl} P.,  et~al., 2010, in Space Telescopes and Instrumentation 2010:
  Ultraviolet to Gamma Ray. p. 77320U (\mn@eprint {arXiv} {1001.2502}),
  \mn@doi{10.1117/12.856577}

\bibitem[\protect\citeauthoryear{{Romer}, {Viana}, {Liddle}  \& {Mann}}{{Romer}
  et~al.}{2001}]{2001ApJ...547..594R}
{Romer} A.~K.,  {Viana} P.~T.~P.,  {Liddle} A.~R.,   {Mann} R.~G.,  2001,
  \mn@doi [\apj] {10.1086/318382}, \href
  {http://adsabs.harvard.edu/abs/2001ApJ...547..594R} {547, 594}

\bibitem[\protect\citeauthoryear{{Rykoff} et~al.,}{{Rykoff}
  et~al.}{2014}]{2014ApJ...785..104R}
{Rykoff} E.~S.,  et~al., 2014, \mn@doi [\apj] {10.1088/0004-637X/785/2/104},
  \href {http://adsabs.harvard.edu/abs/2014ApJ...785..104R} {785, 104}

\bibitem[\protect\citeauthoryear{{Shen}, {Greene}, {Strauss}, {Richards}  \&
  {Schneider}}{{Shen} et~al.}{2008}]{2008Shen}
{Shen} Y.,  {Greene} J.~E.,  {Strauss} M.~A.,  {Richards} G.~T.,   {Schneider}
  D.~P.,  2008, \mn@doi [\apj] {10.1086/587475}, \href
  {http://adsabs.harvard.edu/abs/2008ApJ...680..169S} {680, 169}

\bibitem[\protect\citeauthoryear{{Sijacki}, {Springel}, {Di Matteo}  \&
  {Hernquist}}{{Sijacki} et~al.}{2007}]{2007MNRAS.380..877S}
{Sijacki} D.,  {Springel} V.,  {Di Matteo} T.,   {Hernquist} L.,  2007, \mn@doi
  [\mnras] {10.1111/j.1365-2966.2007.12153.x}, \href
  {http://adsabs.harvard.edu/abs/2007MNRAS.380..877S} {380, 877}

\bibitem[\protect\citeauthoryear{{Steinhardt} \& {Elvis}}{{Steinhardt} \&
  {Elvis}}{2010}]{2010MNRAS.402.2637S}
{Steinhardt} C.~L.,  {Elvis} M.,  2010, \mn@doi [\mnras]
  {10.1111/j.1365-2966.2009.16084.x}, \href
  {http://adsabs.harvard.edu/abs/2010MNRAS.402.2637S} {402, 2637}

\bibitem[\protect\citeauthoryear{{Tie} et~al.,}{{Tie}
  et~al.}{2016}]{2016arXiv161105456T}
{Tie} S.~S.,  et~al., 2016, preprint, \href
  {http://adsabs.harvard.edu/abs/2016arXiv161105456T} {} (\mn@eprint {arXiv}
  {1611.05456})

\bibitem[\protect\citeauthoryear{{Trujillo}, {Graham}  \& {Caon}}{{Trujillo}
  et~al.}{2001}]{2001MNRAS.326..869T}
{Trujillo} I.,  {Graham} A.~W.,   {Caon} N.,  2001, \mn@doi [\mnras]
  {10.1046/j.1365-8711.2001.04471.x}, \href
  {http://adsabs.harvard.edu/abs/2001MNRAS.326..869T} {326, 869}

\bibitem[\protect\citeauthoryear{{Uttley}, {McHardy}  \& {Papadakis}}{{Uttley}
  et~al.}{2002}]{2002MNRAS.332..231U}
{Uttley} P.,  {McHardy} I.~M.,   {Papadakis} I.~E.,  2002, \mn@doi [\mnras]
  {10.1046/j.1365-8711.2002.05298.x}, \href
  {http://adsabs.harvard.edu/abs/2002MNRAS.332..231U} {332, 231}

\bibitem[\protect\citeauthoryear{{Vaughan}, {Fabian}  \& {Nandra}}{{Vaughan}
  et~al.}{2003a}]{Vaughan2003}
{Vaughan} S.,  {Fabian} A.~C.,   {Nandra} K.,  2003a, \mn@doi [\mnras]
  {10.1046/j.1365-8711.2003.06285.x}, \href
  {http://adsabs.harvard.edu/abs/2003MNRAS.339.1237V} {339, 1237}

\bibitem[\protect\citeauthoryear{{Vaughan}, {Edelson}, {Warwick}  \&
  {Uttley}}{{Vaughan} et~al.}{2003b}]{2003MNRAS.345.1271V}
{Vaughan} S.,  {Edelson} R.,  {Warwick} R.~S.,   {Uttley} P.,  2003b, \mn@doi
  [\mnras] {10.1046/j.1365-2966.2003.07042.x}, \href
  {http://adsabs.harvard.edu/abs/2003MNRAS.345.1271V} {345, 1271}

\bibitem[\protect\citeauthoryear{{Vestergaard}}{{Vestergaard}}{2002}]{2002ApJ...571..733V}
{Vestergaard} M.,  2002, \mn@doi [\apj] {10.1086/340045}, \href
  {http://adsabs.harvard.edu/abs/2002ApJ...571..733V} {571, 733}

\bibitem[\protect\citeauthoryear{{Wenger} et~al.,}{{Wenger}
  et~al.}{2000}]{2000A&AS..143....9W}
{Wenger} M.,  et~al., 2000, \mn@doi [\aaps] {10.1051/aas:2000332}, \href
  {http://adsabs.harvard.edu/abs/2000A%26AS..143....9W} {143, 9}

\bibitem[\protect\citeauthoryear{{Woo} \& {Urry}}{{Woo} \&
  {Urry}}{2002}]{2002ApJ...579..530W}
{Woo} J.-H.,  {Urry} C.~M.,  2002, \mn@doi [\apj] {10.1086/342878}, \href
  {http://adsabs.harvard.edu/abs/2002ApJ...579..530W} {579, 530}

\bibitem[\protect\citeauthoryear{{Zhou}, {Zhang}, {Wang}  \& {Zhu}}{{Zhou}
  et~al.}{2010}]{2010ApJ...710...16Z}
{Zhou} X.-L.,  {Zhang} S.-N.,  {Wang} D.-X.,   {Zhu} L.,  2010, \mn@doi [\apj]
  {10.1088/0004-637X/710/1/16}, \href
  {http://adsabs.harvard.edu/abs/2010ApJ...710...16Z} {710, 16}

\bibitem[\protect\citeauthoryear{{van der Klis}}{{van der
  Klis}}{1997}]{1997scma.conf..321V}
{van der Klis} M.,  1997, in {Babu} G.~J.,  {Feigelson} E.~D.,  eds,
  Statistical Challenges in Modern Astronomy II. p.~321 (\mn@eprint {}
  {astro-ph/9704273})

\makeatother
\end{thebibliography}


\newpage
\appendix
\section{Methodology Tests}
\label{MethTests}
We carried out several tests to explore the robustness of the results presented in Section \ref{Scaling}.
\subsection{Luminosity}
\begin{figure}
	\includegraphics[width=\columnwidth]{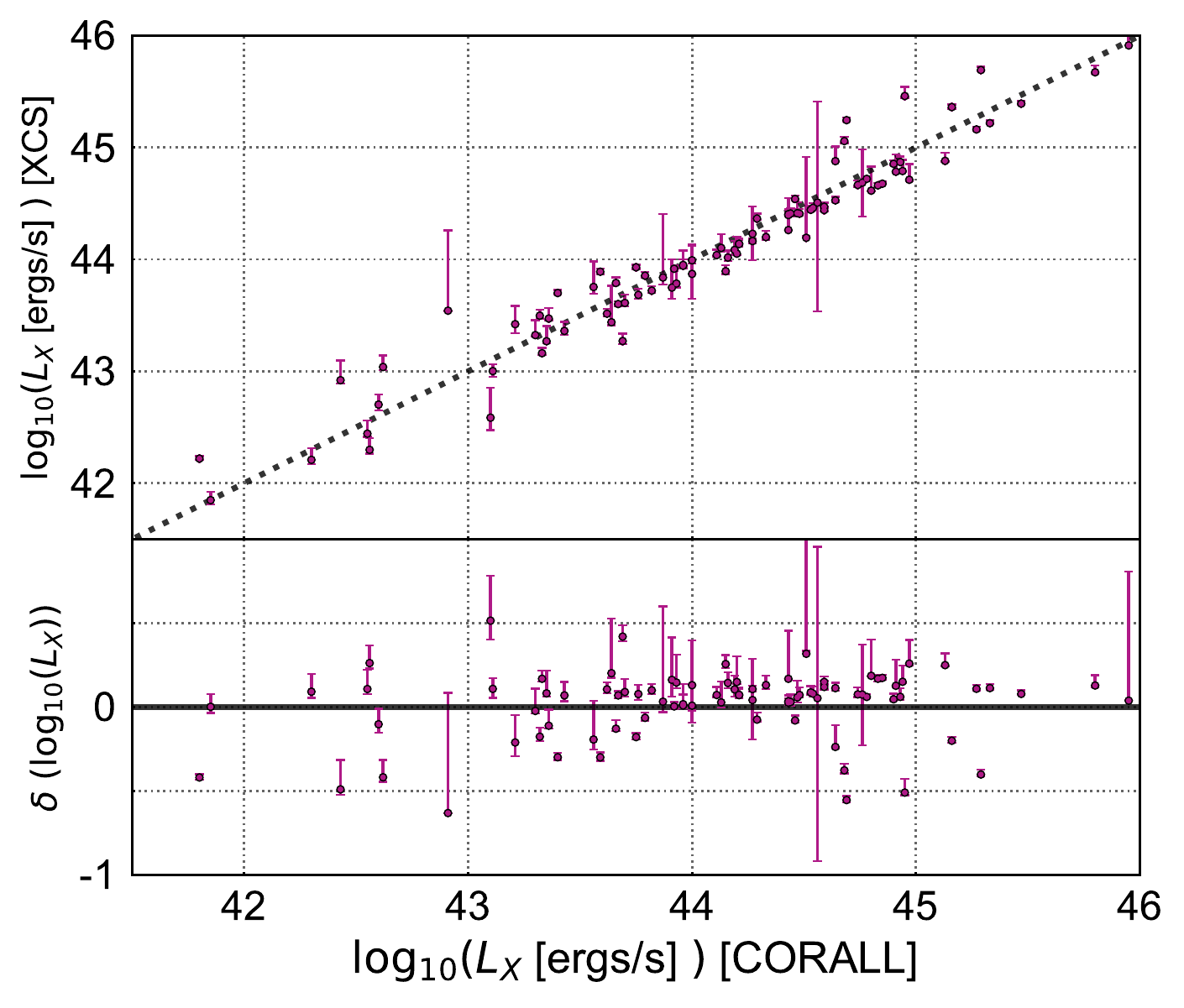}
	\caption{Hard-band (2.0-10\,keV) luminosity of 78 AGN in Sample-S1 common with \citep{2011A&A...530A..42C}. The dotted line is a one-to-one relation (x-axis error bars are not shown for the Corral $L_{\rm{X}}$ estimates as these are not included in their table of results).}
    \label{fig:XCS_v_CorrallSample_hardband_lums}
\end{figure}

We compared our hard-band luminosity results with those estimated by \cite{2011A&A...530A..42C}, a X-ray spectral analysis of $>300$ AGN ($z<2.4$) belonging to the XMM-Newton bright survey (XBS). There are 78 AGN in common with our S1 sample. We found good agreement between the two sets of $L_{\rm{X}}$ measurements, see Figure\, \ref{fig:XCS_v_CorrallSample_hardband_lums}.

\subsection{Normalised excess variance}
\begin{figure}
	\includegraphics[width=\columnwidth]{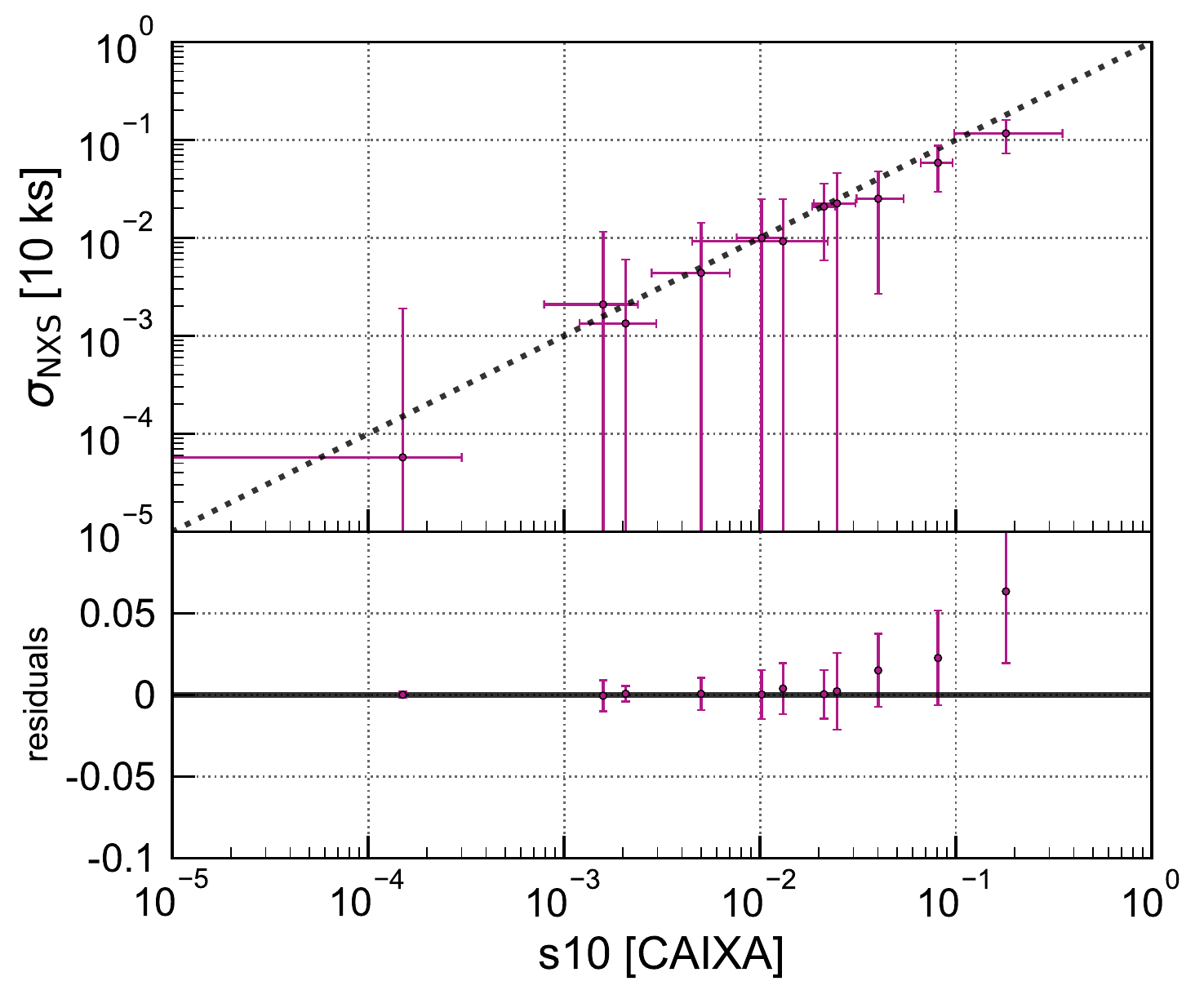}
    \caption{Comparison between $\sigma^2_{\rm{NXS}}$ values derived from 10ks light-curve segments in \citep{Ponti2012}  ($\tt{s10}$ CAIXA ) and our S10 sample. The dotted line is a one-to-one relation.}  
    \label{fig:CAIXAsigma20_v_XCSnxs20kssegs}
\end{figure}

We compared our values for $\sigma^2_{\rm{NXS}}$ with those of the \cite{Ponti2012} CAIXA survey. There are 98 AGN in common with our S1 sample. Of these, there are 12 AGN in the S10 samples (i.e. with 20 or more good 10\,ks light-curve segments). We found a good agreement between our $\sigma^2_{\rm{NXS}}$[20ks] results and the equivalent $\tt{s20}$ values from \citep{Ponti2012}, see Figure\,\ref{fig:CAIXAsigma20_v_XCSnxs20kssegs}. 

As previously noted \citealt{2013ApJ...771....9A} suggest that $\sigma^2_{\rm{NXS}}$ is a biased estimate of variance (even in a continuously sampled light-curve) and provide a scaling factor based on the PSD of the light-curve. Since we don't have prior knowledge of the PSD in most cases, we are unable to use this method -  but should this become available this would be an interesting future approach to determining $\sigma^2_{\rm{NXS}}$.

\subsection{Luminosity contamination by line-of-sight clusters}

\begin{figure}
	\includegraphics[width=\columnwidth]{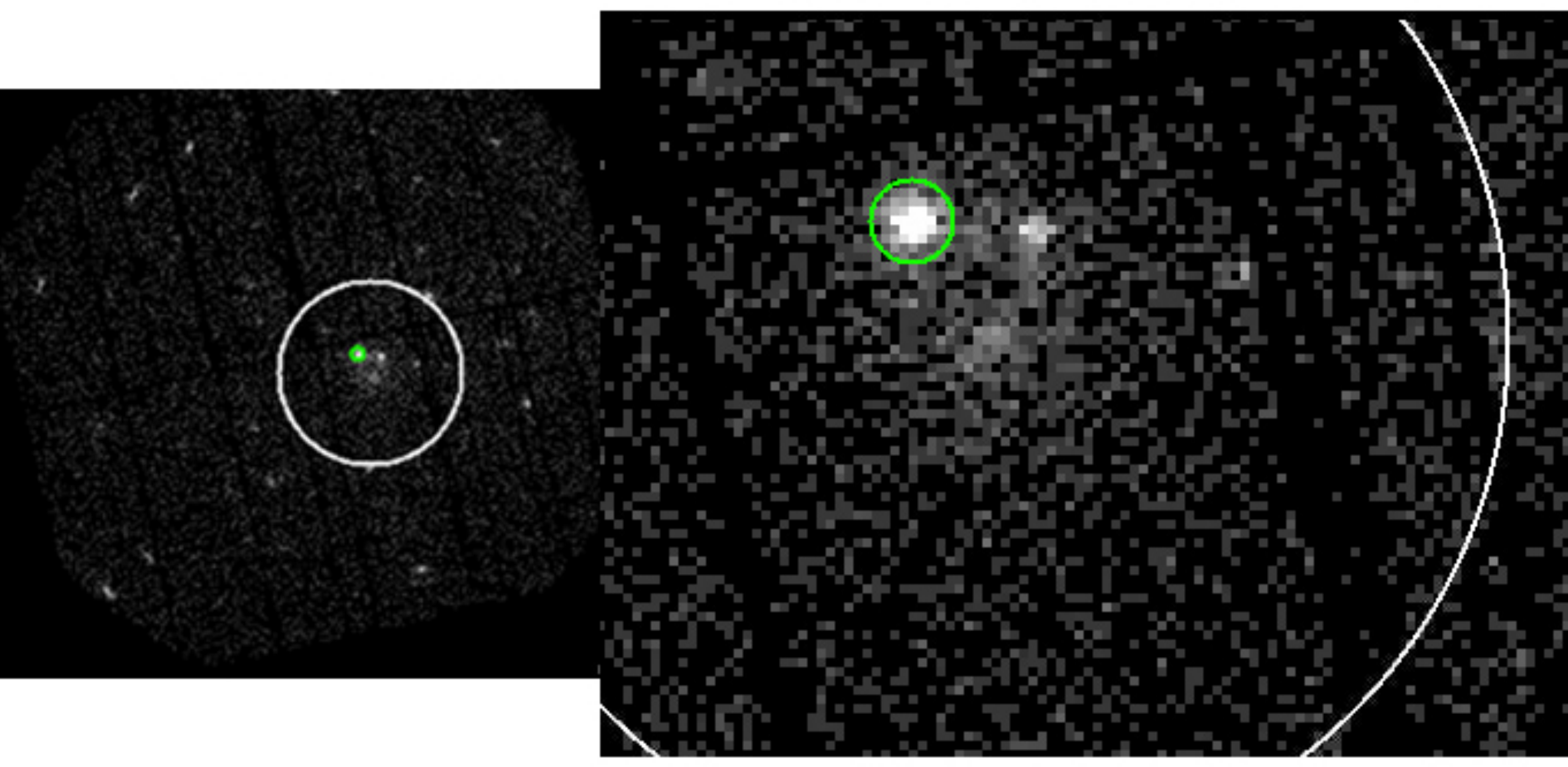}
    \caption{$\tt{PN}$ observation 0305920401 and zoomed image of AGN XMMXCSJ143450.6+033842.5 (green circle). This AGN at $z=0.028$ is within the projected 250\,kpc core radius (white circle) of a $\tt{redMaPPer}$ cluster with a separation of 63\,arcsec. The cluster is at $z=0.146$ at which redshift the angular extent of the core region is 97\,arcsec. The pixel size $4.52''$ in energy range $0.5-2.0\rm{\,keV}$.}
    \label{fig:XMMXCSJ143450}
\end{figure}

We checked to see if any of our AGN lies within or along the line-of-sight to a galaxy cluster. If this was the case, emission from the cluster might boost the measured $L_{\rm{X}}$ of the AGN.  We cross matched all the AGN positions in our S0-Sample with extended XCS sources that  been identified as a cluster in the $\tt{redMaPPer}$ SDSS DR8 cluster catalogue \citep{2014ApJ...785..104R}. For this we used a matching radius of $\sim$250\,kpc, assuming the redshift in the $\tt{redMaPPer}$ catalogue. We found 12 matches (i.e $\sim$1 per cent of sample S1), one of which is shown in Figure\,\ref{fig:XMMXCSJ143450}. None of these 12 were included in the scaling relations presented in \S\,\ref{Scaling} as in all cases there were fewer than five \textit{good} light-curve segments.

\newpage
\section{Tables}
\label{app:AGN_sampleS10_20_40}
\onecolumn
\begin{landscape}
\LTcapwidth=\linewidth
\begin{footnotesize}
\begin{table*}
\begin{center}
   \caption[]{\small{Combined table of Samples S10, S20 and all AGN with $M_{\rm{BH}}$ from reverb mapping in sample S1.   (1) XCS point source name, (2) AGN name if given from literature (3) AGN Type if given from VC13 or SIMBAD, 1, 2 and intermediate, Q-Quasar, BL-BL Lac (4) redshift (5) Luminosity (in 2-10\,keV band) erg\,s$^{-1}$ (6) photon spectral index (7) S10 sample $\sigma^2_{\rm{NXS}}$ value  (8) number of 10\,ks  light-curve segments in AGN,(9) S20 sample $\sigma^2_{\rm{NXS}}$, (10) number of 20\,ks light-curve segments in AGN sample, (11) Black Hole mass M$_{\odot}$ from \citep{Bentz2015}. }\label{tab:A3}}
\begin{tabular}{ccccccccccc}
\hline
 XCS Name & AGN Name & Type & $z$ & log($L_{\rm{X}}$) & $\Gamma$ & $\sigma^2_{\rm{NXS10ks}}$ & $N_{10{\rm{ks}}}$ & $\sigma^2_{\rm{NXS20 ks}}$ & $N_{20{\rm{ks}}}$& log$(M_{\rm{BH}})$  \\ [0.5ex]
 &&&&${\rm{erg\,s^{-1}}}$&&&&&&$M_\odot$ \\
 (1)&(2)&(3)&(4)&(5)&(6)&(7)&(8)&(9)&(10)&(11) \\
  \hline
  \hline
XMMXCS J000619.5+201210.3&MARK  335&1&0.026&42.840$\pm^{0.003}_{0.002}$&1.52&0.01$\pm{0.015}$&35&&&7.23$\pm^{0.042}_{0.044}$\\
XMMXCS J001030.9+105829.6&III Zw 2&1.2&0.089&44.126$\pm^{0.026}_{0.019}$&1.427&&&&&8.067$\pm^{0.119}_{0.165}$ \\
XMMXCS J005452.3+252538.1&PG0052+251&1.2&0.154&44.580$\pm^{0.007}_{0.006}$&2.002&&&&&8.462$\pm^{0.083}_{0.094}$ \\
XMMXCS J012345.6-584822.5&F9&1.2&0.047&43.825$\pm^{0.003}_{0.003}$&1.846&&&&&8.299$\pm^{0.078}_{0.116}$ \\
XMMXCS J021433.4-004601.4&MARK  590&1.0&0.026&42.889$\pm^{0.014}_{0.012}$&1.686&&&&&7.570$\pm^{0.062}_{0.074}$ \\
XMMXCS J033336.4-360826.1&NGC 1365&1.8&0.006&41.2$\pm^{0.036}_{0.034}$&1.07&0.007$\pm{0.01}$&67&0.008$\pm{0.015}$&36&\\
XMMXCS J043311.0+052116.1&3C120&1.5&0.033&44.924$\pm^{0.003}_{0.003}$&1.526&&&&&7.745$\pm^{0.038}_{0.040}$ \\
XMMXCS J051045.4+162956.7&2E 0507+1626&1.5&0.018&43.196$\pm^{0.003}_{0.003}$&1.495&&&&&6.876$\pm^{0.074}_{0.424}$  \\
XMMXCS J051611.4-000859.5&AKN  120&1.0&0.033&44.568$\pm^{0.001}_{0.001}$&1.961&&&&&8.068$\pm^{0.048}_{0.063}$ \\
XMMXCS J055947.3-502652.2&PKS 0558-504&1&0.137&45.57$\pm^{0.002}_{0.002}$&2.3&0.005$\pm{0.01}$&37&&&\\
XMMXCS J074232.7+494833.3&MARK   79&1.2&0.022&42.919$\pm^{0.006}_{0.006}$&1.096&&&&&7.612$\pm^{0.107}_{0.136}$ \\
XMMXCS J081058.6+760243.0&PG 0804+761&1.0&0.100&44.173$\pm^{0.006}_{0.005}$&2.361&&&&&8.735$\pm^{0.048}_{0.053}$ \\
XMMXCS J084742.5+344504.0&PG 0844+349&1.0&0.064&42.977$\pm^{0.016}_{0.016}$&1.153&&&&&7.858$\pm^{0.154}_{0.230}$ \\
XMMXCS J092512.7+521711.7&MARK  110&1n&0.035&43.873$\pm^{0.004}_{0.003}$&1.920&&&&&7.292$\pm^{0.101}_{0.097}$ \\
XMMXCS J095651.9+411519.7&PG 0953+415&1.0&0.234&44.688$\pm^{0.009}_{0.007}$&2.196&&&&&8.333$\pm^{0.082}_{0.108}$ \\
XMMXCS J102348.6+040553.7&ACIS J10212+0421&1&0.099&42.17$\pm^{0.135}_{0.009}$&3.85&0.007$\pm{0.05}$&21&&&\\
XMMXCS J103118.3+505333.9&1ES 1028+511&BL&0.361&45.42$\pm^{0.002}_{0.002}$&2.32&0.0$\pm{0.007}$&23&&&\\
XMMXCS J103438.6+393825.8&KUG 1031+398&1&0.042&42.12$\pm^{0.011}_{0.012}$&2.48&0.01$\pm{0.015}$&27&&&\\
XMMXCS J110647.4+723407.0&NGC 3516&1.5&0.009&42.45$\pm^{0.003}_{0.003}$&0.85&0.002$\pm{0.01}$&35&&&7.395$\pm^{0.037}_{0.061}$\\
XMMXCS J112916.6-042407.6&MARK 1298&1&0.06&42.59$\pm^{0.022}_{0.023}$&&0.01$\pm{0.028}$&22&&&\\
XMMXCS J120309.5+443153.0&NGC 4051&1&0.002&41.16$\pm^{0.003}_{0.003}$&1.53&0.058$\pm{0.029}$&41&0.068$\pm{0.037}$&23&6.13$\pm^{0.121}_{0.155}$\\
XMMXCS J121417.6+140313.9&PG 1211+143&1&0.082&43.72$\pm^{0.003}_{0.003}$&1.78&0.004$\pm{0.01}$&34&&&\\
XMMXCS J121826.5+294847.1&MARK  766&1&0.013&42.62$\pm^{0.003}_{0.003}$&1.77&0.021$\pm{0.015}$&47&0.023$\pm{0.02}$&27&6.822$\pm^{0.05}_{0.057}$\\
XMMXCS J122324.2+024044.9&MARK   50&1.2&0.023&43.011$\pm^{0.005}_{0.004}$&1.830&&&&&7.422$\pm^{0.057}_{0.068}$ \\
XMMXCS J122548.8+333249.0&NGC 4395&1.8&0.001&40.838$\pm^{0.010}_{0.009}$&0.595&&&&&5.449$\pm^{0.130}_{0.145}$ \\
XMMXCS J122906.6+020309.0&3C 273.0&1.0&0.158&46.003$\pm^{0.001}_{0.001}$&1.670&&&&&8.839$\pm^{0.077}_{0.113}$ \\
XMMXCS  J123203.7+200928.1&TON 1542&1.0&0.063&43.445$\pm^{0.016}_{0.013}$&2.144&&&&&7.758$\pm^{0.175}_{0.219}$ \\
XMMXCS J123939.4-052043.3&NGC 4593&1.0&0.009&42.811$\pm^{0.004}_{0.003}$&1.848&&&&&6.882$\pm^{0.084}_{0.104}$ \\
XMMXCS J130022.1+282402.8&X COM&1.5&0.092&43.57$\pm^{0.005}_{0.004}$&2.33&&&0.001$\pm{0.01}$&22&\\
XMMXCS J132519.2-382455.2&IRAS 13224-3809&1&0.065&42.66$\pm^{0.009}_{0.008}$&2.53&0.117$\pm{0.044}$&42&0.157$\pm{0.051}$&25&\\
XMMXCS J133553.7-341745.5&MCG -06.30.015&1.5&0.008&42.55$\pm^{0.004}_{0.004}$&1.6&0.023$\pm{0.023}$&27&&&\\
XMMXCS J134208.4+353916.1&NGC 5273&1.9&0.004&41.127$\pm^{0.008}_{0.007}$&0.909&&&&&6.660$\pm^{0.125}_{0.190}$ \\
XMMXCS J135303.7+691828.9&MARK  279&1.0&0.030&43.513$\pm^{0.003}_{0.003}$&1.946&&&&&7.435$\pm^{0.099}_{0.133}$ \\
XMMXCS J141759.5+250812.2&NGC 5548&1.5&0.017&43.4$\pm^{0.003}_{0.002}$&1.8&0.001$\pm{0.005}$&74&0.0$\pm{0.001}$&32&7.718$\pm^{0.016}_{0.016}$\\
XMMXCS J153552.2+575411.7&MARK  290&1.5&0.030&43.231$\pm^{0.006}_{0.006}$&1.535&&&&&7.277$\pm^{0.061}_{0.061}$ \\
XMMXCS J155543.0+111125.4&PG 1553+11&BL&0.36&45.42$\pm^{0.001}_{0.001}$&2.28&0$\pm{0.002}$&77&0.002$\pm{0.003}$&38&\\
XMMXCS J172819.6-141555.7&PDS 456&Q&0.184&44.45$\pm^{0.003}_{0.002}$&2.26&0.001$\pm{0.005}$&71&0.004$\pm{0.01}$&34&\\
XMMXCS J190525.8+422739.8&Zw 229.015&1&0.028&42.803$\pm^{0.006}_{0.005}$&1.782&&&&&6.913$\pm^{0.075}_{0.119}$ \\
XMMXCS J194240.5-101924.5&NGC 6814&1.5&0.005&42.090$\pm^{0.003}_{0.003}$&1.680&&&&&7.038$\pm^{0.056}_{0.058}$ \\
XMMXCS J204409.7-104325.8&MARK 509&1.5&0.035&44.03$\pm^{0.002}_{0.002}$&2.13&0$\pm{0.002}$&62&0$\pm{0.003}$&31&8.049$\pm^{0.035}_{0.035}$\\
XMMXCS J213227.8+100819.6&UGC 11763&1.5&0.063&43.472$\pm^{0.020}_{0.016}$&1.691&&&&&7.433$\pm^{0.055}_{0.063}$ \\
XMMXCS J215852.0-301332.4&PKS 2155-304&BL&0.116&44.75$\pm^{0.003}_{0.002}$&2.65&&&0.001$\pm{0.006}$&23&\\
XMMXCS J224239.3+294331.9&AKN  564&2&0.025&43.89$\pm^{0.002}_{0.002}$&2.55&0.025$\pm{0.023}$&29&&&\\
XMMXCS J230315.6+085223.9&NGC 7469&1.5&0.016&43.247$\pm^{0.002}_{0.003}$&1.974&&&&&6.956$\pm^{0.048}_{0.050}$ \\
  \hline
  
\end{tabular}
\end{center}
\end{table*}
\end{footnotesize}
\end{landscape}


\bsp	
\label{lastpage}
\end{document}